\RequirePackage{fix-cm}
\documentclass[twocolumn, natbib]{svjour3}          % twocolumn

\smartqed  % flush right qed marks, e.g. at end of proof

\usepackage{graphics}
\usepackage{epsfig}
\usepackage{subfigure}
\usepackage{units}
\usepackage[intlimits]{amsmath}
\usepackage[normalem]{ulem}
\usepackage{color}
\usepackage{ulem}
\journalname{Experiments in Fluids}

\begin{document}
%%%%%%%%%%%%%%%%
%%%%%%%%%%%%%%%%

\title{Scalar dissipation rate measurements in a starting jet%\thanks{Grants or other notes
}


\author{N. Soulopoulos \and%First Author         \and
        Y. Hardalupas  \and%Second Author %etc.
        A.M.K.P. Taylor
}


\institute{N. Soulopoulos \at
           Mechanical Engineering Department, \\
           Imperial College London, \\
           Exhibition Road, SW7 2BX, London, UK \\
           \email{ns6@ic.ac.uk}           %  \\
           \and
           Y. Hardalupas \at
           Mechanical Engineering Department, \\
           Imperial College London, \\
           \and
           A.M.K.P. Taylor \at
           Mechanical Engineering Department, \\
           Imperial College London, \\
}

\date{Received: date / Accepted: date}

\maketitle

\begin{abstract}
Measurements of the scalar dissipation rate are performed in an impulsively started gas jet, using planar laser induced fluorescence. The measurements are well resolved spatially. The deteriorating effect of experimental noise on this experiment is treated with a Wiener filter, which is shown to be applicable to this large-scale inhomogeneous flow. The accuracy of the scalar dissipation rate is within 20\%, as determined from an explicit calculation of the filtering errors. The residual fields that remain after the filtering are analysed in detail and their statistical properties show that these resemble white noise to a good approximation. The level of corrections is minimal for the scalar field but it is of the order of 40\% for the scalar dissipation rate. An examination of the filtering operation using modeled spectra and the measured spatial resolution shows that the Wiener filter produces errors in the estimate of the scalar dissipation rate $\sim30\%$, for Taylor-scale Reynolds number up to 1000. The implications of this modelling are discussed with respect to common experimental situations and point out the relative merits of improving the spatial resolution as compared to improvements in the signal to noise ratio.
% Include keywords, PACS and mathematical subject classification numbers as needed.
%\PACS{PACS code1 \and PACS code2 \and more}
\end{abstract}

%%%%%%%%%%%%%%%%%%%%%%
\section{Introduction} \label{section_introduction}

The dissipation rate of scalar fluctuations in a turbulent flow, $\chi$, is defined as
\begin{equation}
\chi=D\left[\left(\partial\zeta/\partial x\right)^2+\left(\partial\zeta/\partial y\right)^2
+\left(\partial\zeta/\partial z\right)^2\right]
\label{eq_scalar_dissipation_rate}
\end{equation}
where $\zeta$ is a scalar and $D$ is its diffusivity. It is of fundamental interest in mixing between fluids, \cite{Sreenivasan91,warhaft_review_2000,Dimotakis05}, and is a measure of the degree of active mixing that is in progress in a turbulent flow. For example, in homogeneous turbulence, the scalar fluctuations are 'smoothed out' by the scalar dissipation rate, \cite{TennekesLumley}, and as a consequence its increase leads to faster molecular mixing between fluids. Considerable attention has been paid to the measurement of the scalar dissipation in variety of flows, which has served to establish basic and common characteristics of this quantity in a variety of flows. For example, \cite{AntoniaMi93,joint_statistics_anselmet_1994} used cold wires to measure the dissipation of temperature fluctuations (considered as a passive scalar) in axially symmetric jets and \cite{AntoniaAnselmet85,AntoniaBrowne83a} used the same technique in plane jets. In \cite{Sreenivasan93}, a variety of flows and techniques were used to identify common characteristics of the scalar dissipation rate in both gaseous and liquid flows. Furthermore, \cite{dowling_estimated_dissipation_1991} and \cite{pitts_sdr_1999} used point and line, respectively, Rayleigh scattering to measure time series of the concentration and its dissipation in round gas jets. The spatial structure of the scalar dissipation field was demonstrated, in liquid and gaseous round jets, by \cite{DahmBuch96} and \cite{DahmBuch98}, respectively. They used planar laser induced fluorescence and planar Rayleigh scattering, respectively, to record 2-dimensional images of instantaneous dissipation structures; measurements of these structures in a gaseous planar jet were performed by \cite{su_clemens_2003}. Additionally, 2-dimensional Raman scattering was used, \cite{schefer_sdr_1988,schefer_sdr_1994}, to measure the scalar dissipation rate in an axisymmetric gaseous jet in order to elucidate the role of large scales on the jet development. Three-dimensional measurements of the scalar dissipation rate have, also, been attempted in gaseous and liquid jets and wakes, for example by \cite{Sreenivasan90a}. Other flows have also received attention, for example grid turbulence, \cite{Warhaft94b}, where passive temperature fluctuations and their dissipation were measured using cold wires. Finally, the scalar dissipation rate in opposed jet flows was measured using either pairs of cold wires, \cite{SardiTaylor98}, or planar laser induced fluorescence of acetone, \cite{Danaila02a}.

Furthermore, in the case of non-premixed flames, chemical reaction happens only after molecular mixing between fuel and oxidiser, so the scalar dissipation rate is of prime importance in combustion as well, where it arises naturally in the expression of the mean reaction rate, \cite{bilger76}. Additionally, in the flamelet modelling approach of non-premixed turbulent flames \cite{williams_flamelet_1975,peters_flamelets_1984} the scalar dissipation rate of the mixture fraction (a normalised fuel concentration) provides the link between chemistry and flow field. There is evidence as well, \textit{e.g.} \cite{mastorakos_dns_autoignition_1997}, that the scalar dissipation rate is a key quantity controlling the autoignition of two, initially separate, gases. Despite their considerable difficulty, measurements of the scalar dissipation rate of mixture fraction or temperature in non-premixed flames have been performed. For example, Rayleigh scattering has been performed to acquire point or 2-dimensional images of temperature and its dissipation in turbulent jet flames, \textit{e.g.} \cite{Dahmetal95,Peters88,frank_dissipation_structures_2007}, and it was, also, combined with planar laser induced fluorescence to measure fuel concentration, \cite{Driscoll02}. Many measurements of the dissipation rate of mixture fraction in turbulent jet diffusion flames have been performed using line Raman scattering, combined with laser induced fluorescence; for example, \cite{Pitz94,Mansour97} measured hydrogen jet flames, whereas \cite{BilgerLong94,BarlowKarpetis02,Janicka05} measured the dissipation rate in hydrocarbon flames.

The scalar dissipation rate is a small scale quantity, so the measurement probe volume should ideally be of the order of the Kolmogorov or Batchelor length scales, \cite{locally_axisymmetric_turbulence_george_1991}. However, at these small resolutions experimental noise can become comparable with the measured signal. Many studies have shown the effect of experimental noise and spatial resolution on the measurement of the scalar dissipation rate using either analytical techniques, computational results or experimental data, for example \cite{Wyngaard71,AntoniaMi93b,mi_nathan_resolution_2003,burattini_dissipation_corrections_2007,ghandhi_resolution_noise_2005}. It is generally true that in ideal, noise-free measurements, the magnitude of the scalar dissipation increases with increasing spatial resolution and converges to the true value for well-resolved measurements. Similarly, the experimental noise tends to increase the magnitude of the scalar dissipation with increasing spatial resolution. However, it is known that as the measurement probe volume becomes very small, this increase is unbounded and faster than the increase due to resolution effects; it is possible to measure many times the actual average value of the dissipation due to noise effects. So, under-resolved, but noisy, measurements can give the 'correct' value of $\left<\chi\right>$. Therefore, measurements of the mean scalar dissipation rate have to be checked, and possibly corrected, for noise and spatial resolution effects and measurements of the instantaneous scalar dissipation require post-processing of the scalar measurement, typically in the form of digital filtering for noise removal.

It is possible to check that the mean scalar dissipation rate has been measured correctly by either using theoretical arguments (mentioned below) or by verifying the balance of the scalar variance transport equation, \textit{e.g.} \cite{AntoniaMi93,Bilger04}. Furthermore, resolution correction factors for the mean scalar dissipation, in homogeneous and isotropic turbulence, rely in averaging a three-dimensional spectrum of the scalar fluctuations, either using a modelled form \cite{Wyngaard68,Wyngaard71}, or using measured one-dimensional spectra \cite{AntoniaMi93b}. These results show that the spatial resolution required to measure the mean scalar dissipation rate within 10\% of the true value is $2\sim3\eta_B$ ($\eta_B$ being the Batchelor length scale), when cold wires are used for the measurement of passive temperature fluctuations.

Various methods have been devised to calculate the mean scalar dissipation rate from measurements of the underlying scalar. In a slightly heated axially symmetric jet \cite{AntoniaMi93} employed a pair of cold wires and used two techniques for the determination of the mean scalar dissipation rate. The measured temperature-derivative spectra were used to devise correction factors for the mean scalar dissipation rate, by adapting the above mentioned analysis \cite{Wyngaard71}; furthermore, a Taylor series expansion of the cross correlation coefficient for small distances was extrapolated to zero separation, giving an estimate of the mean dissipation rate. The two techniques agreed very well and the accuracy of the mean scalar dissipation rate was determined by showing that it balances the scalar variance transport equation. The cross correlation coefficient technique can also be used for non-homogeneous flows by including higher order terms in the Taylor series expansion, \cite{AntoniaKrishna}. A method that relies on the second order structure function estimates the noise variance by plotting the mean squared scalar difference for various separations. An extrapolation of this plot to zero separation gives a measure of the noise variance, as was done in a heated round jet \cite{joint_statistics_anselmet_1994} and in a heated opposed jet flow \cite{SardiThesis}, using pairs of cold-wires. Then, the second order structure function, minus the noise variance, is divided by the squared separation and plotted as a function of the separation. A final extrapolation to zero separation gives the mean scalar dissipation rate. Although the above methods measure accurately the mean scalar dissipation rate, it is always desirable to ensure a high signal-to-noise ratio. It should be noted that these methods are mainly applicable to homogeneous and isotropic turbulence.

Other techniques, for the measurement of the mean scalar dissipation rate, rely on over-sampling or redundant scalar measurements. For example, \cite{frank_dissipation_structures_2007} measured the temperature signal in a turbulent non-premixed flame using 2-dimensional laser Rayleigh scattering. The fact that noise is uncorrelated between adjacent rows (or columns) of the image data allows the calculation of power spectra by interlacing between consecutive rows. The noise level was reduced in the scalar derivative spectra, from which the mean scalar dissipation can be calculated. Also, in experiments in a turbulent jet flame using single-point laser Rayleigh scattering \cite{ClemensVarghese} used either over-sampling or two different probe volumes to measure the temperature fluctuations at the same spatial location. The noise variance was then identified and subtracted from the measurement of the mean scalar dissipation and corrections of up to 90\% of the uncorrected dissipation were obtained. Finally, the kinetic energy dissipation rate was calculated at two different resolutions from a single set of data \cite{eaton_2007_dissipation_correction_piv} and, by combining the two measures, the correct mean dissipation rate was inferred (this technique can be readily adapted to the measurement of the scalar dissipation rate). Since these techniques do not rely on the assumptions of homogeneity and isotropy, they seem broader in scope for the determination of the mean scalar dissipation rate. However, they do not provide accuracy estimates for the resulting measured mean dissipation rate (the method in \cite{eaton_2007_dissipation_correction_piv} does provide a measure of its accuracy).

The measurement of the instantaneous scalar dissipation rate is more restrictive in terms of signal-to-noise ratio, while being equally important. So, it generally requires the application of digital filtering, or other post-processing, to the raw scalar measurements. In measuring the passive temperature fluctuations in a jet flow using cold wires, \cite{nathan_mi_filtering} used an iterative procedure to filter the raw data, by continuously adjusting the cut-off frequency of a digital filter. The algorithm used Kolmogorov scalings and converged to the digital cut-off frequency when this value and the updated Kolmogorov frequency were the same within a very small bound. In jet flame experiments \cite{frank_dissipation_structures_2007} obtained the instantaneous scalar dissipation rate by filtering the instantaneous 2D temperature images using gaussian kernels, whose widths were different at each of the two spatial directions and, also, varied with the position in the flow. Recently, \cite{tong_2009_conditional_sampling_noise_reduction_sdr} proposed a method that relied, also, on the measured scalar field and the expectation that the so-called sub-grid scale variance of the mixture fraction is small at well-resolved, low-noise portions of the data. In this way, the noise variance could be obtained and used to correct under-resolved data. In other experiments, \cite{dimotakis_scalar_spectra_water_jet_1996,Danaila02a} used an optimal Wiener filter approach, that is also adopted in the present experiment, to filter Laser Induced Fluorescence (LIF) measurements in a round jet and in a partially stirred reactor, respectively. Finally, \cite{mastorakos_sdr_2005} used a filtering method that relies on a wavelet decomposition of the raw scalar measurements.

The above exposition shows that in the case of homogeneous and isotropic turbulence, the signal processing involved in measuring the mean scalar dissipation rate can be checked against theoretical considerations to ascertain the accuracy of the measurement. In a more general flow configuration, concerning also the measurement of the instantaneous scalar dissipation rate, the signal processing takes a central role in determining the scalar dissipation rate, so it is important to fully analyse the post processing procedure. For example, filters usually rest on certain assumptions about the underlying signal and it is indeed possible to check the consistency of the filtering operation. This is necessary because the effects of filtering, for example the level of corrections, the residual fields and their characteristics, are not usually demonstrated for any given filtering operation. Furthermore, many of the existing methods for correcting the mean scalar dissipation rate do not provide corrections for the instantaneous scalar fields.

In the present paper we develop an approach to address the issues above by using the Wiener filtering theory \cite{dimotakis_scalar_spectra_water_jet_1996}. We show how to explicitly calculate the errors that arise in the filtering process and so give an accuracy estimate for the mean scalar dissipation rate. We, also, present in detail the effects of filtering on the measured scalar fields. It is important to note that the Wiener filter provides both a noise and a resolution correction (since it makes use of the Point Spread Function of the measuring system) for the instantaneous scalar fields. We show that the Wiener filter is rather insensitive to the modelling involved, while at the same time is easily applied; it is, however, a 'global' method, since it relies on scalar power spectra. In order to verify that the noise and the resolution corrections are accurate, we compare the mean scalar dissipation obtained from the Wiener filter with results from another technique \cite{eaton_2007_dissipation_correction_piv}, which has been shown to correct for noise and resolution effects as well as being second order accurate in the spatial resolution (thus giving an estimate of the accuracy). However, we choose the Wiener filter because we can filter instantaneous images and obtain the distributions of the scalar dissipation rate and its unconditional and conditional (on the scalar value) statistics.

We use a transient, incompressible jet flow in our measurements, since this flow is of technological importance, \textit{e.g.} gaseous fuel injection in engines, and, also, has received limited attention in terms of mixing measurements. It is more common to report velocity measurements in starting jet experiments in order to calculate the very important parameter of ambient air entrainment. For example, \cite{cossali_unsteady_jet_velocity_2001} used Laser Doppler Anemometry to map the velocity field in a large region of a transient jet having Reynolds number Re\unit[$\sim12*10^{3}$]{} and \cite{witze_starting_jet_1980} performed hot -film anemometer velocity measurements on the centreline of a starting jet at Re\unit[$\sim7*10^{3}$]{}. Also, computational studies of gaseous starting jets have been performed for similar Reynolds numbers, \textit{e.g.} \cite{abraham_near_field_entrainment_2000}. In particular, \cite{entrainment_starting_jet_LES_2012} calculated the starting jet experiment by \cite{witze_starting_jet_1980} in order to gain further understanding of the behaviour of transient jets as related to internal combustion engines. However, the relevant Reynolds number in, for example, high pressure diesel spray injection depends on the prevailing conditions in the fuel nozzle and in the combustion chamber. Assumptions for characteristic quantities give a Reynolds number $\sim15*10^3$, \textit{e.g.} \cite{gosman_2004_primary_atomization_diesel}, or $\sim150*10^3$ when using state methodologies for the calculation of viscosity at supercritical conditions in modern engines, \cite{diesel_spray_les_oefelein_2012}.

The structure of the paper is as follows. Section \ref{section_exp_arrangement} describes the flow field and the optical measurement technique. Section \ref{section_digital_filtering} describes the Wiener filter along with a calculation of the ensuing filtering errors and presents a comparison with another dissipation correction method \cite{eaton_2007_dissipation_correction_piv}. Section \ref{section_results} provides an assessment of the effect of filtering and demonstrates scalar dissipation rate measurements in a starting jet flow. Finally, section \ref{section_conclusions} gives a summary and conclusions.

%%%%%%%%%%%%%%%%%%%%%%%%%%%%%%%%%%
\section{Experimental arrangement} \label{section_exp_arrangement}

This section describes the flow configuration, the optical arrangement for laser induced fluorescence measurements and gives details for the measurement spatial resolution in relation to the flow length scales. The flow is that of an unsteady gaseous jet inside a low velocity co-flow. A starting jet is created by suddenly releasing fluid from a nozzle. It has received relatively little attention compared with the steady state jet. However, it appears in a variety of important engineering applications. For example, the injection inside the combustion chamber of a compression-ignition internal combustion engine is generally short, so that the fuel jet does not reach a steady state. In such engines, the mixing afforded by the injection event is crucial in controlling emissions.

\subsection{Flow configuration} \label{section_flow}
\begin{figure}[htbp]
\centering
\includegraphics[scale=0.4]{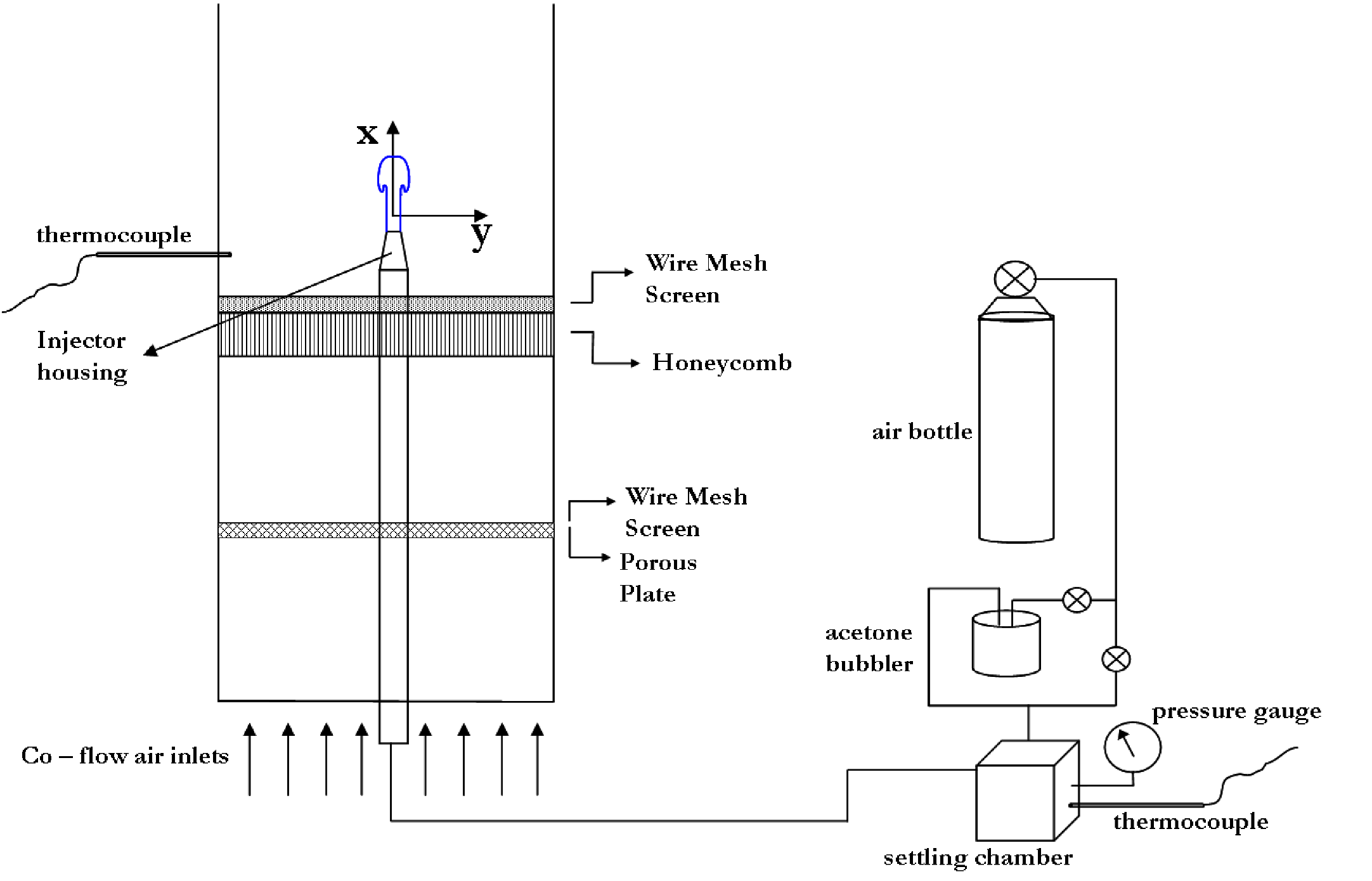}
\caption{The experimental arrangement and the coordinate system used.}
\label{fig_flow_chamber}
\end{figure}

The flow is established inside a square chamber of 30$*$30 \unit[]{cm$^{2}$} cross section. A general view of the flow chamber and of the coordinate system used is shown in figure \ref{fig_flow_chamber}. The bottom part of the chamber serves as the co-flow conditioning area and the upper part, which is open to the atmosphere, is the injection and measurement area.

The co-flowing air, provided by a compressor, is heated using inline air process heaters, settles in a manifold and feeds the flow chamber through 8 inlets. The average cross-sectional temperature of the air inside the flow chamber is \unit[65]{$^{\circ}$C}$\pm$\unit[7]{$^{\circ}$C}, in order to prevent acetone condensation, up to a downstream distance of \unit[100]{mm}, equal to 25 nozzle diameters, from the jet nozzle exit. The temperature during data collection remained practically constant and the above variation consisted of differences in temperature across the chamber, with the temperature near the chamber walls being slightly smaller. The jet injection and the optical measurements take place after a flow conditioning section that ensures a low velocity (around \unitfrac[0.1]{m}{s}), low turbulence, homogeneous velocity profile. Due to the low velocity of the co-flow, we estimate that air entrainment inside the injector nozzle, in-between injections, is minimal.

The injector is a commercial, automotive gas injector from KEIHIN (model KN3-2B). It consists of a fast acting solenoid valve, which is placed \unit[20]{mm} upstream of the nozzle exit, and a straight tube of $d$=\unit[4]{mm} internal diameter, with a lip thickness of about 1 mm. It is connected to a settling chamber which, in turn, is connected to a pressurised air bottle through a pressure regulator that keeps the pressure inside the settling chamber constant. A pressure gauge and a K-type thermocouple are used to monitor the pressure and the temperature inside the settling chamber, upstream of the injector. The back pressure of the injection inside the settling chamber is \unit[2.5]{bar}.

The jet air flow passes, first, through inline process heaters and, later, through an acetone bubbler before being injected. The acetone bubbler is kept inside a water bath at \unit[50]{$^{\circ}$C} in order to maximise the acetone signal and keep the level of seeded acetone constant during a run, whereas the temperature of the air stream downstream of the bubbler and just before the injector is kept at \unit[65]{$^{\circ}$C}. The co-flow air starts to flow well in advance of any measurements, so that thermal equilibrium between the injector nozzle and the co-flow air is achieved. The concentration of acetone in the jet air is such that there is no discernible absorption of the laser light along the diameter of the jet, at distances very near the nozzle exit.

In order to identify the temporal response of the injector's solenoid valve opening, the acetone seeded fluid exiting the injector is imaged and the time delay between the solenoid valve activation signal and the first camera signal is determined. Then, the "travel" time of the fluid in the straight pipe section between the valve and the nozzle exit is subtracted and the resulting time is considered as representing the solenoid valve opening time. The ensuing time is less than \unit[1]{ms}.

The injection velocity is determined from the calibration curve provided by the manufacturer of the injector and gives the output flow rate as a function of back pressure (pressure inside the settling chamber) and injection duration. In the absence of detailed velocity measurements, the injection velocity profile as a function of time (the injection schedule) is assumed to be a top hat profile, \textit{i.e.} the fluid velocity is constant throughout the injection event and this velocity is reached instantaneously. The injection duration is \unit[10]{ms} and the first appearance of jet fluid at the nozzle exit happens at about \unit[5]{ms} after the initiation of the injection pulse to the injector driver.

The injection velocity is $u$=\unitfrac[20]{m}{s} and the Reynolds number, assuming a kinematic viscosity $\nu$=\unitfrac[1.8$*$10$^{-5}$]{m$^{2}$}{s}, is Re=\unit[4500]{}. From the velocity and the nozzle diameter a time scale is calculated as $T=d/u$=\unit[0.2]{ms}.

During every injection, one image is acquired at a given time After the Start of Injection (ASI) and at every time ASI 300-500 images are acquired, in order to collect time-dependent, ensemble-averaged measurements. Injections are taking place every \unit[500]{ms}, so as to avoid interaction between two consecutive injections and prevent acetone condensation inside the air lines. Images are acquired every 4th injection.

\subsection{Optical measurement} \label{section_optical_measurement}
\begin{figure}[htb]
\centering
\includegraphics[scale=0.55]{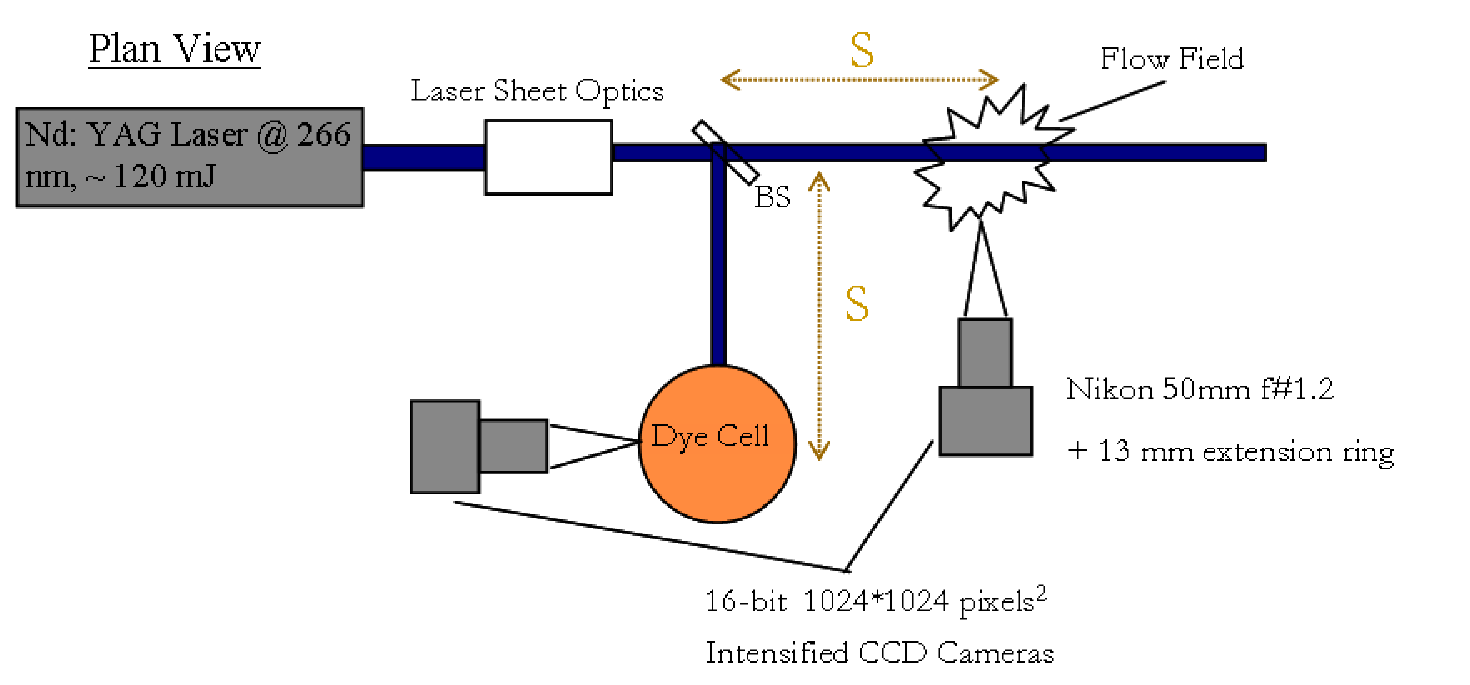}
\caption{Plan view of the optical setup.}
\label{fig_optical_setup_plan_view}
\end{figure}

The 4th harmonic of a pulsed, Nd:YAG laser at \unit[266]{nm} (\unit[10]{ns} pulse duration) is used to excite the acetone molecules and a 16-bit Intensified CCD camera (Andor DH534-18F-03) is used to record the fluorescent intensity. A BG-3 (Schott) colour filter rejects light at the laser wavelength and allows the fluorescent light at the blue region of the spectrum to pass through. Three cylindrical fused silica lenses with focal lengths \unit[-20]{mm}, \unit[104]{mm} and \unit[310]{mm} are used to generate a \unit[45]{mm} high and \unit[130]{$\mu$m} thick laser sheet. The camera is fitted with a \unit[50]{mm}, f\#1.2 (f\# is the lens f-number) Nikkon lens and with a \unit[13]{mm} extension ring for tighter focusing. Before entering the flow chamber, a beam-splitter reflects 5\% of the laser sheet energy into a dye cell containing a solution of Rhodamine 6G. A camera captures the emitted fluorescence signal, so that the laser energy and the laser sheet profile are monitored on a shot-by-shot basis. The laser energy at the probe volume is \unit[100]{mJ/pulse}. A plan view of the optical setup is given in figure \ref{fig_optical_setup_plan_view}.

For conditions of constant temperature and pressure and at the optically thin limit, the acetone fluorescent intensity is proportional to, and varies linearly with, the laser energy and the acetone concentration. The images from the dye cell are used to correct the acquired acetone LIF images for the laser sheet profile and energy; typically, the inter-shot variation of the laser energy was of the order of 2\% and, after these corrections, the spatial variation of the fluorescent signal along a radial profile at the exit of the nozzle was of the order of 2\%; the camera non-linearities were minimal and were not corrected for. The fluorescent signal can further be processed to provide quantitative measurements of the mixture fraction. In the present two-fluid system the mixture fraction is the contribution of the jet fluid stream to the local concentration. The corrected LIF signal intensity at any given spatial position in the flow is $I(x,y)$ and the acetone LIF signal within the potential core at the nozzle exit is $I_{0}$. This intensity corresponds to pure jet fluid, so the mixture fraction is given at every point in the flow as $\zeta(x,y)=I(x,y)/I_{0}$, where $\zeta$ is the mixture fraction. In all images, the field of view includes the nozzle exit, so the acetone LIF signal corresponding to jet fluid is always known. The 2-dimensional approximation to the scalar dissipation rate is obtained from the mixture fraction field by using second order finite differences to calculate the spatial derivatives, for example, $\partial\zeta/\partial x=\left(\zeta_{i+1,j}-\zeta_{i-1,j}\right)/\left(2\Delta x\right)$, where $\Delta x$ is the pixel size and $\zeta_{i+1,j}=\zeta\left(x+\Delta x,y\right)$. the choice of the differentiation scheme depends on the resolution and signal-to-noise ratio of the measurements. While a higher order differentiation scheme has better spectral response, its outcome depends on the relative length scales between the scalar structure and the spatial resolution. We, also, used a scheme that rotates the pixel array by 45 degrees and calculates the derivative as an average of the original and rotated fields. The resulting difference in the mean scalar dissipation rate is less than 2\% from the values obtained through 2nd order differencing.

\subsection{Spatial resolution and flow length scales} \label{section_spatial_resolution}
The spatial resolution of the measurement depends on the laser sheet thickness and the in-plane spatial resolution. The in-plane spatial resolution is not necessarily the same as the pixel spacing (which is \unitfrac[0.045]{mm}{pixel} for the 1024$*$1024 pixels$^{2}$ CCD array) due to the blurring effect of the imaging system (camera lens, pixel size and intensifier). Using standard optics methods \cite{siegman_book}, \cite{eckbreth}, the laser sheet thickness is estimated to be \unit[130]{$\mu$m}, at the $1/e^{2}$ position of the gaussian laser beam, growing to \unit[150]{$\mu$m} at the edges of the imaged region -- we also measured the laser beam thickness, using a scanning knife edge method, and found that it was $\sim$\unit[150]{$\mu$m}.

\begin{figure}[htb]
\centering
\includegraphics[scale=0.4]{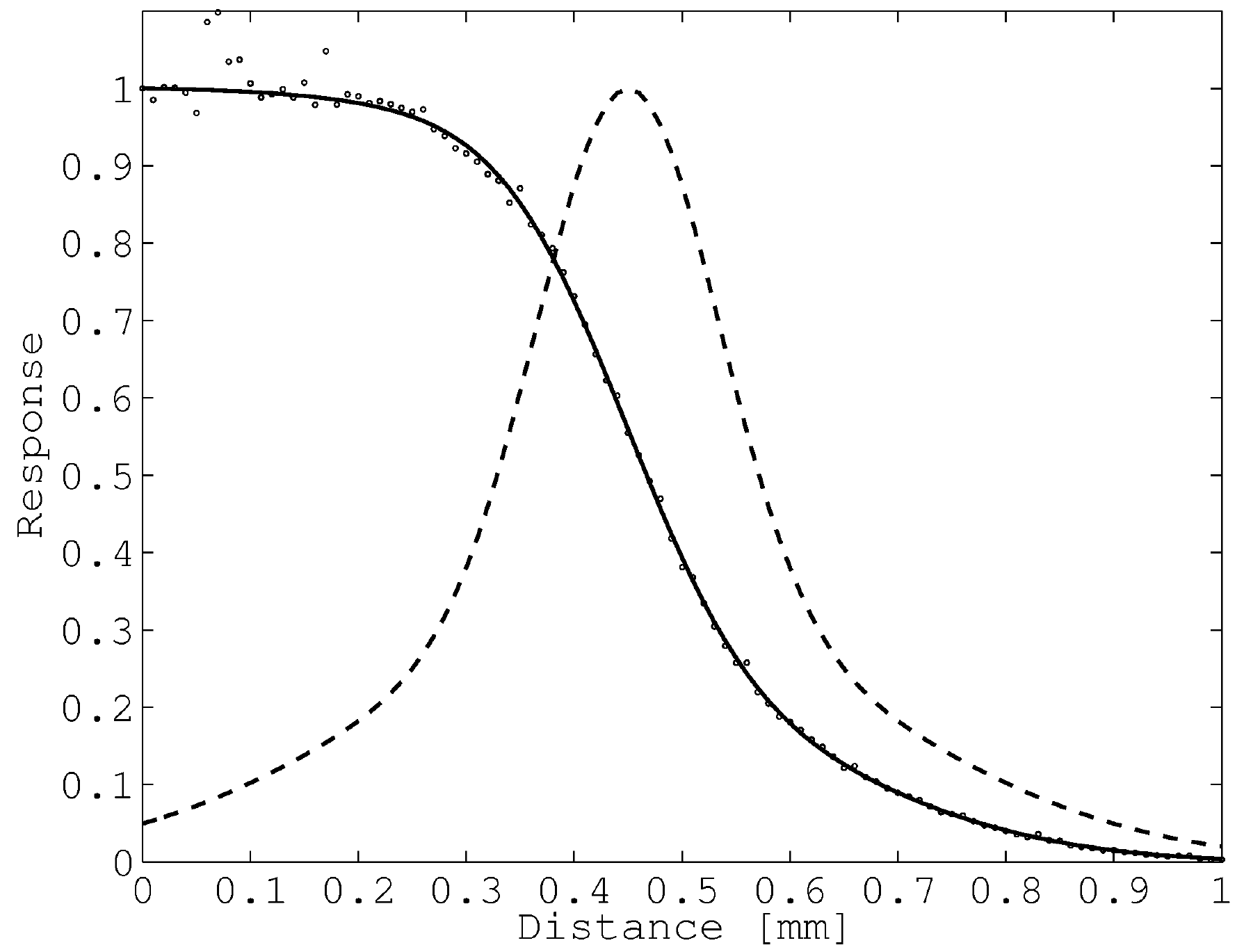}
\caption{Measurement of the optical system resolution. The symbols correspond to the measured intensity distribution at a pixel as a function of the knife edge position along the CCD array and the solid line is a spline fit to the data points. The derivative of the fit is the dashed line, known as the line spread function.}
\label{fig_psf_estimate}
\end{figure}

The blurring of the optical system is characterised by the Point Spread Function (PSF) (and its Fourier transform, the Modulation Tranfer Function - MTF) \cite{goodman_fourier_optics_book}, which quantifies the spreading of intensity values over the pixels when imaging a point source. The recorded intensity, $f_{M}(x,y)$, of an imaged object, $f(x,y)$, is given by $f_{M}(x,y)=\int_{-\infty}^{+\infty}\!h(x-x',y-y')f(x',y')\,\mathrm{d}x'\mathrm{d}y'$, where $h(x,y)$ is the PSF. In the case where the resolution of the optical system is limited by the camera lens, the PSF would be expected to have a form similar to gaussian. Where an image intensifier is employed, the form of the PSF might have longer tails than a gaussian. We measured the resolution of the optical system using the scanning knife edge technique. Figure \ref{fig_psf_estimate} shows the measured data points along with a spline fit and the derivative of the fit, the so-called line spread function. The Fourier transform of the line spread function is the MTF along a given direction, so it is closely related to the PSF. The resulting full width at half maximum of the LSF is \unit[0.24]{mm}, however the LSF shows "tails" that are longer than an equivalent gaussian profile, which is sometimes used for fitting the LSF.

The smallest length scale of variation of the velocity field in a turbulent flow is the Kolmogorov length scale defined as $\eta=\left(\nu^{3}/\varepsilon\right)^{1/4}$, where $\nu$ is the kinematic viscosity of the fluid and $\varepsilon$ is the kinetic energy dissipation rate.

With respect to a passive scalar field the smallest scale of motion depends on the relative magnitudes of momentum and mass diffusivity, as given by the Schmidt number, $Sc=\nu/D$ ($D$ is the molecular diffusivity). The smallest length scale of the scalar field is then given by the Batchelor length scale as $\eta_{B}=\eta Sc^{-1/2}$, \cite{batchelor_small_scale_temperature_1}, \cite{batchelor_small_scale_temperature_2}, \cite{CorrsinSpectrum}. When Sc$\sim$1, as in the present mixing of two incompressible gases, the Kolmogorov and Batchelor length scales are the same.

\begin{figure}[thb]
\centering
\begin{tabular}{c}
\includegraphics[scale=0.4]{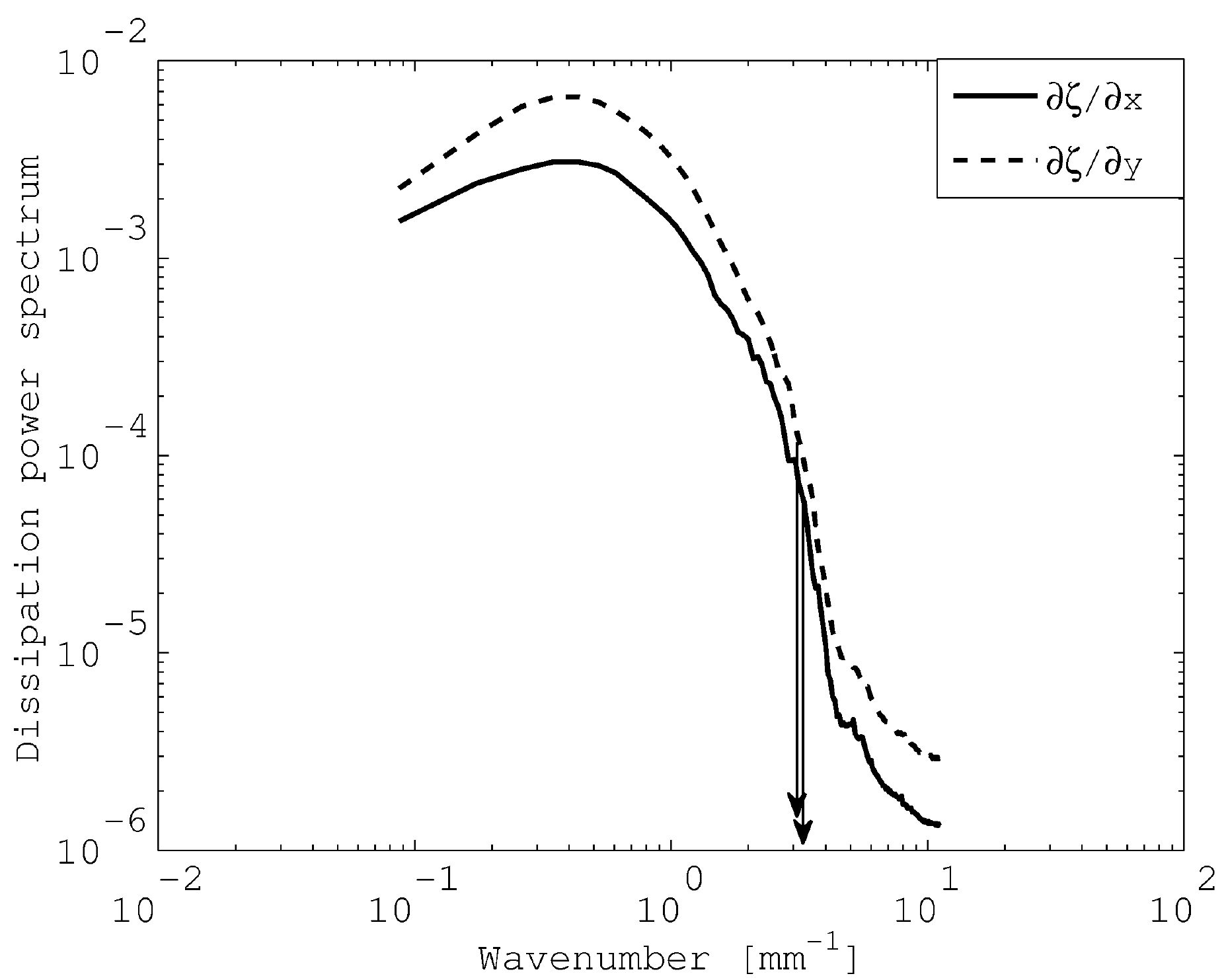}
\end{tabular}
\caption{Determination of the Batchelor length scale from the measured dissipation power spectra. The vertical arrows show the wavenumber corresponding to 2\% of the peak value.}
\label{fig_resolution_spectra_t10}
\end{figure}

A way to estimate the Batchelor length scale of the flow is to use the scalar energy spectrum \cite{dissipation_scales_wang_barlow}. According to a model spectrum \cite{Pope_book} the wavenumber, $\kappa$, where $\kappa\eta=1$ corresponds to a value of the dissipation spectrum equal to 2\% of the peak value. The dissipation spectra are calculated in the following way: a small square region (around 50 pixels on each side) at a position in the flow and at a given time ASI is chosen and the spectra of the mixture fraction axial gradient for all columns in the region, and for all images, are calculated and averaged together so as to obtain an ensemble-averaged spectrum of $\partial\zeta/\partial x$ in the axial direction; the same happens for all rows of the region, corresponding to the spectrum of $\partial\zeta/\partial y$ in the radial direction. Before calculating the spectrum, a 2D Hann window is applied to the derivative data. This helps to minimise edge effects that arise from the abrupt cutoff of the signal as a result of the windowing of the data. Figure \ref{fig_resolution_spectra_t10} shows these spectra for a flow region at the spatial location of the average position of the core of the vortex ring at time t/T=13.4 ASI and, also, shows the wavenumber corresponding to 2\% of the peak value of the spectrum. For the axial-derivative spectra the wavenumber is $\sim\unit[3.3]{mm^{-1}}$ that corresponds to an estimated Batchelor scale of $\sim\unit[300]{\mu\mathrm{m}}$ and for the radial-derivative spectra the corresponding numbers are $\sim\unit[3.1]{mm^{-1}}$ and $\sim\unit[320]{\mu\mathrm{m}}$, respectively. For other times ASI and other flow positions the values range from $\sim\unit[250]{\mu\mathrm{m}}$ to $\sim\unit[320]{\mu\mathrm{m}}$. So, taking as an estimate of the Batchelor length scale the smaller of the two, $\eta_{B}\sim \unit[250]{\mu\mathrm{m}}$. So, these estimates confirm that the probe volume size offers appropriate spatial resolution for the measurements.

\begin{figure}[htb]
\centering
\includegraphics[scale=0.4]{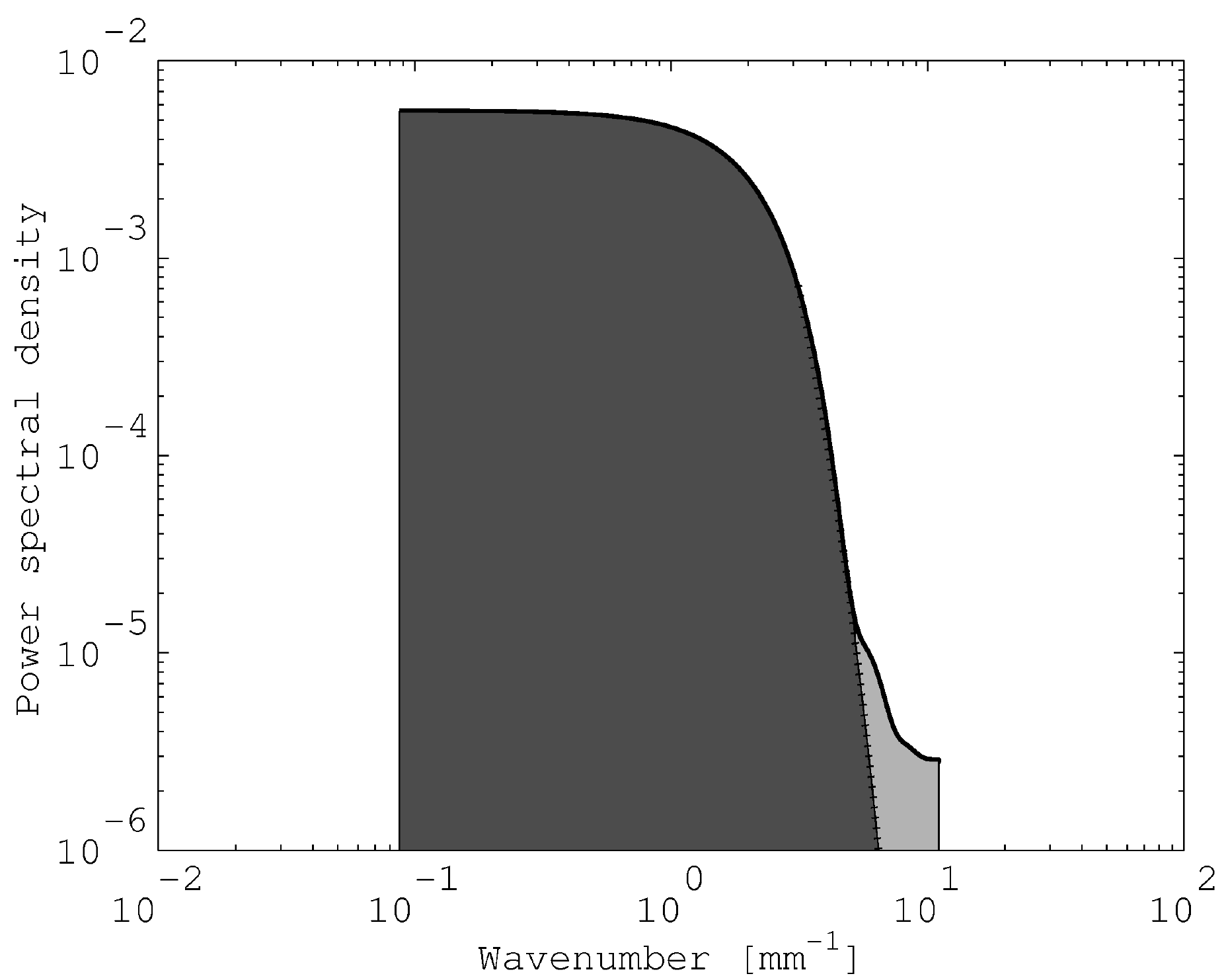}
\caption{The spectrum of the scalar fluctuations for a small region in the flow centred at (x/D,y/D)=(4,1.15), showing the noise contribution (light filled area) and the signal, with an estimated high-wavenumber variation (dark filled area). The ratio of those two areas defines the signal-to-noise ratio of the measurements, at this particular location.}
\label{fig_noise_spectrum_explain}
\end{figure}

Finally, the signal-to-noise ratio of the measurements can be deduced using the scalar energy power spectra calculated above. Exponential functions are fitted at the high wavenumber portion of the spectra which permit the calculation of the ratio of the integral of the fitted spectrum over the integral of the measured minus the fitted spectra. These areas are shown in figure \ref{fig_noise_spectrum_explain} as the darker and lighter shaded areas, respectively. This ratio is the signal-to-noise ratio (SNR), which was always SNR$>$50 and typically SNR$\approx$100, depending on the position in the flow and the time after the start of injection. Following the same procedure for the scalar gradient power spectra results in SNR$\approx$10, \textit{i.e} an order of magnitude lower than in the case of the mixture fraction. These numbers show that the mixture fraction measurements can be considered of good quality, however still the determination of the scalar dissipation rate is problematic.

%%%%%%%%%%%%%%%%%%%%%%%%%%%
\section{Digital filtering} \label{section_digital_filtering}

This section presents the theory behind the Wiener filter and shows how to calculate the remaining errors, which can be considered as an estimate of the accuracy of the scalar dissipation rate measurements. Finally, the Wiener filter is compared with an analytically-derived method which corrects for noise and resolution effects and gives an estimate of the accuracy, in order to verify the validity of the assumptions used in the Wiener filter.

\subsection{Optimal Wiener filter} \label{section_optimal_wiener_filter}
In the following discussion, lowercase letters denote intensity images in physical space and uppercase letters the
corresponding images in wavenumber space. The coordinates in physical space are $r=(x,y)$ and the corresponding symbols in wavenumber space are $\kappa=(u,v)$.

An image blurred by a Point Spread Function (PSF) and corrupted by additive noise can be written as
\begin{equation}
f_{M}(r)=\int_{-\infty}^{+\infty}\!h(r-r')f(r')\,\mathrm{d}r' + n(r)
%\Leftrightarrow F_{M}(\kappa)=H(\kappa)F(\kappa)+N(\kappa)
\label{eq_image_model}
\end{equation}
or $F_{M}(\kappa)=H(\kappa)F(\kappa)+N(\kappa)$, where $f_{M}(r)$ is the measured image, $h(r)$ is the PSF of the measuring system, $f(r)$ is the noise-free, unblurred image, and $n(r)$ is the zero-mean noise. Using the above measurement model, the Wiener filter tries to reconstruct the ideal image, $f(r)$, by applying a filter $g(r)$ to the measured image $f_{M}(r)$. Then, the filtered image can be written as
\begin{equation}
f_{F}(r)=\int_{-\infty}^{+\infty}\!g(r-r')f_{M}(r')\,\mathrm{d}r'
%\Leftrightarrow F_{F}(\kappa)=G(\kappa)F_{M}(\kappa)
\label{eq_wiener_estimate}
\end{equation}
or $F_{F}(\kappa)=G(\kappa)F_{M}(\kappa)$. The filtering error is defined as $e(r)=f(r)-f_{F}(r)$ and the mean square error is
\begin{equation}
e^{2}=E\left[\left(f(r)-f_{F}(r)\right)^{2}\right]
\label{eq_wiener_mean_square_error}
\end{equation}
where $E$ denotes the expectation. The Wiener filter is the optimum solution that minimises this mean square error. A complete exposition of this minimisation problem is given in \cite{petrou_fundamentals}, where it is proven that the Fourier transform of the function $g(r)$ is $G(\kappa) = W(\kappa)/H(\kappa)$, where
\begin{equation}
W(\kappa) = \frac{|H(\kappa)|^{2}}{|H(\kappa)|^{2}+\frac{S_{n}(\kappa)}{S(\kappa)}}
\label{eq_wiener_filter}
\end{equation}
is the Fourier transform of the Wiener filter. The power spectral densities of the noise and of the uncorrupted, noise-free image are $S_{n}(\kappa)=|N(\kappa)|^{2}$ and $S(\kappa)=|F(\kappa)|^{2}$, respectively and the filtered image is
\begin{equation}
F_{F}(\kappa)=\frac{W(\kappa)F_{M}(\kappa)}{H(\kappa)}
\label{eq_wiener_filtered_image}
\end{equation}

A characteristic of the Wiener filter is that it requires knowledge of the power spectral densities of both the unblurred, noise-free image and the noise. A possibility, followed by \cite{Danaila02a} in the context of fluid mechanics and, also, suggested by \cite{petrou_fundamentals} for a more general situation, is the following: a series of measured images can be used to plot an estimate of the spectral density of the signal plus noise. With increasing wavenumbers the power spectral density of the signal decreases, up to the point where the noise starts to have a significant contribution. Then, the spectrum of the true signal, for the highest wavenumbers, can be modelled by a decreasing exponential function. Such a high-wavenumber form of the three dimensional spectrum has been proposed at various model spectra, \cite{CorrsinSpectrum,pao_spectrum,batchelor_small_scale_temperature_2,Pope_book}. Similarly, a suitable fit (\textit{e.g.} using cubic splines) can be found for the noise density and extrapolated to the wavenumbers where the true signal dominates the power spectrum.

\begin{figure}[htbp]
\centering
\includegraphics[scale=0.5]{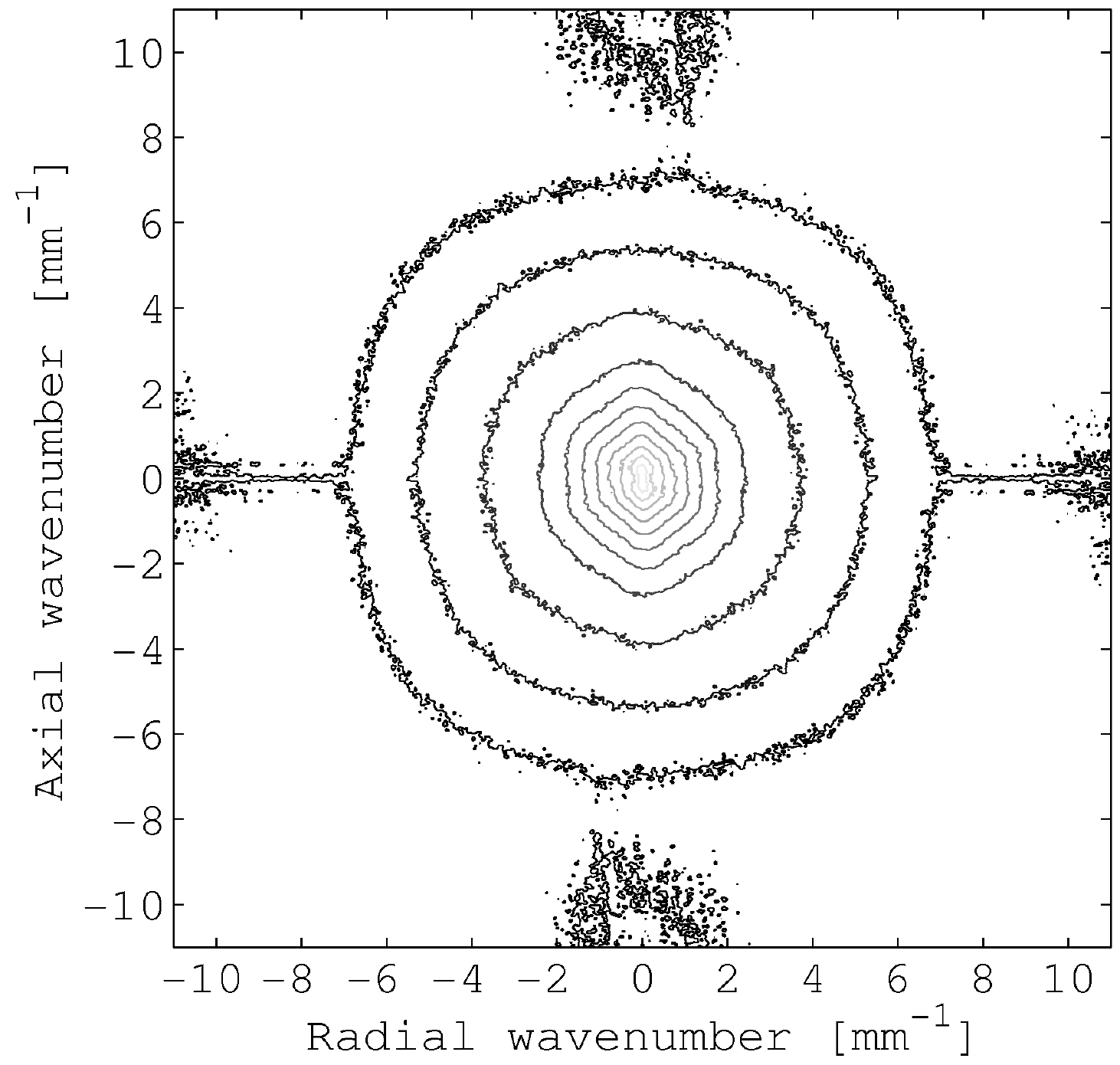}
\caption{An averaged 2D spectrum of the mixture fraction fluctuations at time t/T=9.65.}
\label{fig_spectrum_2D}
\end{figure}

\begin{figure}[htbp]
\centering
\includegraphics[scale=0.4]{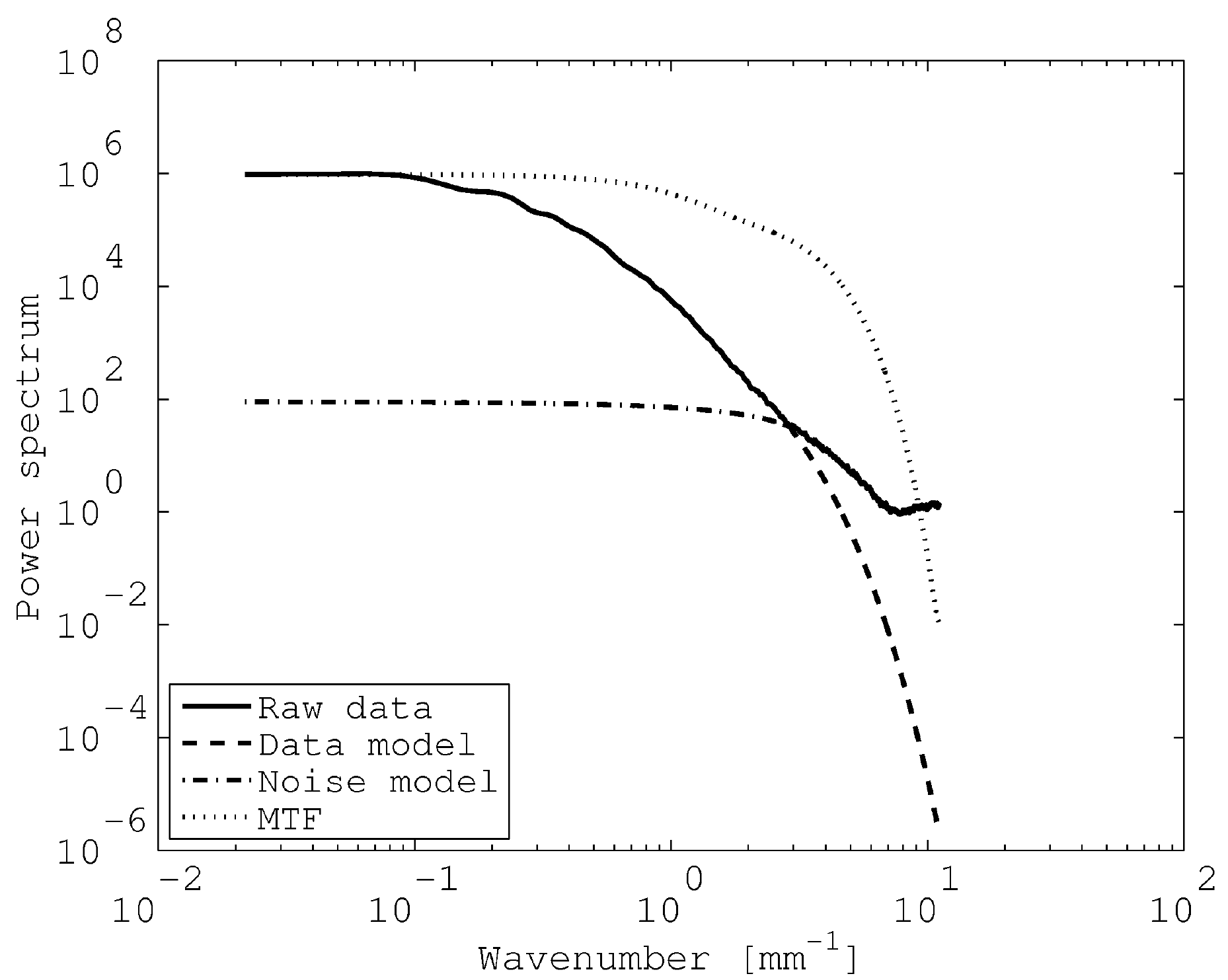}
\caption{A horizontal cut through the averaged 2D spectrum of the mixture fraction fluctuations at time t/T=9.65. The model signal spectrum for the high wavenumbers, the model spectrum for the noise and the modulation transfer function (which is the Fourier transform of the PSF) are also shown.}
\label{fig_wiener_model_spectra}
\end{figure}

Figure \ref{fig_spectrum_2D} shows an ensemble-averaged 2D spectrum of the mixture fraction fluctuations at time t/T=9.65. At the low wavenumbers the contours of the 2D spectra are not circular which demonstrates the in-homogeneity of the flow field at the large scales of motion. Nevertheless, at higher wavenumbers the contours become circular and the exponential fitting only takes place after this radial symmetry has been established. The models used for the construction of the Wiener filter at t/T=9.65 are shown in figure \ref{fig_wiener_model_spectra}, which plots a horizontal cut through the centre of the frequency plane and along the radial direction of the averaged 2D spectrum of figure \ref{fig_spectrum_2D}, alongside the model spectra of the high wavenumber region of the signal and of the noise. The portion of the spectrum that is modelled in the determination of the Wiener filter will be discussed in section \ref{section_wiener_method_noise}.

\begin{figure}[htbp]
\centering
\includegraphics[scale=0.4]{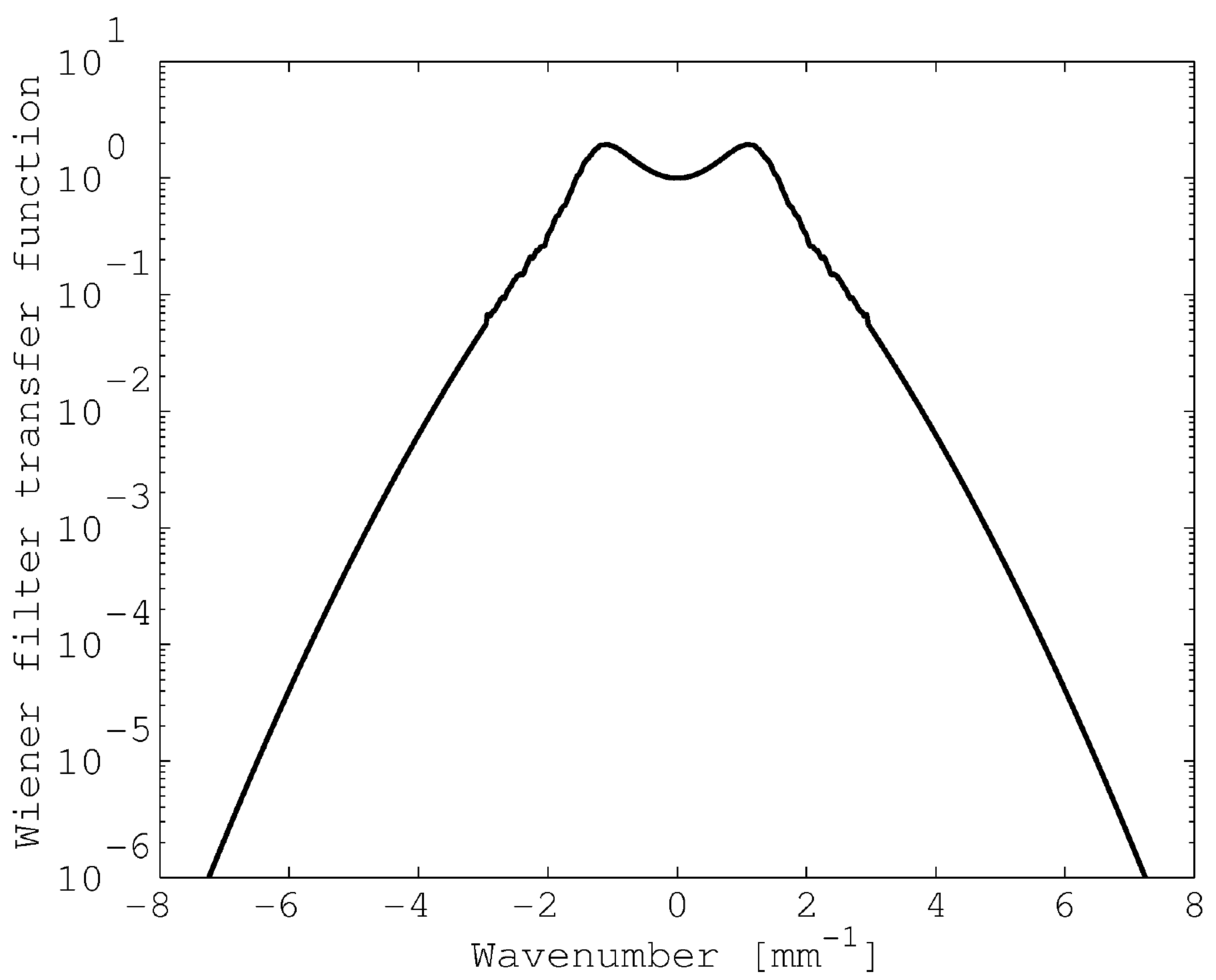}
\caption{The transfer function of the Wiener filter at time t/T=9.65.}
\label{fig_wiener_filter_T08}
\end{figure}

The Wiener filter calculated from Eq.\ref{eq_wiener_filter} at time t/T=9.65 is shown in figure \ref{fig_wiener_filter_T08} and it shows a "bump", as a characteristic of the Wiener filter with the inclusion of the PSF.

\subsection{Filtering errors} \label{section_filtering_errors}
The derivation of the Wiener filter provides the possibility to calculate the ensuing filtering errors for the variance of the mixture fraction and the scalar dissipation fields. By using Parseval's theorem, a different way to express the mean square error, Eq. \ref{eq_wiener_mean_square_error}, is
\begin{equation}
e^{2}=\int_{-\infty}^{+\infty}\!|f(r)-f_{F}(r)|^{2}\,\mathrm{d}r
     =\int_{-\infty}^{+\infty}\!|F(\kappa)-F_{F}(\kappa)|^{2}\,\mathrm{d}\kappa
\label{eq_wiener_mean_square_error_freq}
\end{equation}
Using Eqs. \ref{eq_wiener_estimate} and \ref{eq_wiener_filtered_image} and taking into account that the signal and the noise are uncorrelated, the mean square error, Eq. \ref{eq_wiener_mean_square_error_freq}, can be written as (omitting the dependence on $\kappa$)
\begin{equation}
e^{2}=\int_{-\infty}^{+\infty}\!\frac{1}{H^{2}}\left[F_{S}^{2}(W-1)^{2}+N^{2}W^{2}\right]\,\mathrm{d}\kappa
\label{eq_wiener_mean_square_error_expand}
\end{equation}
where $F_{S}=HF$. If the optimum filter from Eq. \ref{eq_wiener_filter} is substituted in the above equation, the mean square error becomes
\begin{equation}
e^{2}=\int_{-\infty}^{+\infty}\!\frac{1}{H^{2}}\frac{F_{S}^{2}N^{2}}{F_{S}^{2}+N^{2}}\,\mathrm{d}\kappa
\label{eq_wiener_error}
\end{equation}
The mean square error in the spatial derivative is calculated by considering the following mean square error form
\begin{equation}
e^{2}_{d}=e^{2}_{d,x}+e^{2}_{d,y}
=\int_{-\infty}^{+\infty}\!
\left[
\left|\frac{\partial f}{\partial x}-\frac{\partial f_{F}}{\partial x}\right|^{2}
+
\left|\frac{\partial f}{\partial y}-\frac{\partial f_{F}}{\partial y}\right|^{2}
\right]
\,\mathrm{d}r
\label{eq_wiener_mean_square_error_deriv_def}
\end{equation}
where
$e^{2}_{d,x}=\int_{-\infty}^{+\infty}\!
\left|\partial f/\partial x-\partial f_{F}/\partial x\right|^{2}\,\mathrm{d}r$
and
\[
e^{2}_{d,x}=
\int_{-\infty}^{+\infty}\!
\left|iuF(\kappa)-iuF_{F}(\kappa)\right|^{2}\,\mathrm{d}\kappa
=
\int_{-\infty}^{+\infty}\!
u^{2}\left|F(\kappa)-F_{F}(\kappa)\right|^{2}\,\mathrm{d}\kappa
\]
and similarly for $e^{2}_{d,y}$. So,
\[
e^{2}_{d}=\int_{-\infty}^{+\infty}\!
\kappa^{2}\left|F(\kappa)-F_{F}(\kappa)\right|^{2}\,\mathrm{d}\kappa
\]
where $\kappa^{2}=u^{2}+v^{2}$. The expression $\left|F(\kappa)-F_{F}(\kappa)\right|^{2}$ appears in the mean square error of Eq. \ref{eq_wiener_mean_square_error_freq} and is given in Eq. \ref{eq_wiener_error}, when the optimal Wiener filter is used. Then, the associated mean square error for the spatial derivative of the mixture fraction field is
\begin{equation}
e^{2}_{d}=\int_{-\infty}^{+\infty}\!
\kappa^{2}\frac{1}{H^{2}}\frac{F_{S}^{2}N^{2}}{F_{S}^{2}+N^{2}}\,\mathrm{d}\kappa
\label{eq_wiener_mean_square_error_deriv}
\end{equation}

The mean square filtering errors were evaluated from Eqs. \ref{eq_wiener_error} and \ref{eq_wiener_mean_square_error_deriv} and are shown in table \ref{table_wiener_errors}, where $\zeta$ is the instantaneous mixture fraction, for all the times after the start of injection. The mean square errors of the mixture fraction field are given as the ratio of the Eq. \ref{eq_wiener_error} and the integral over all wavenumbers of the model spectrum, $\int_{-\infty}^{+\infty}\!S(\kappa)\,\mathrm{d}\kappa$, \textit{i.e.} the "true" variance. The corresponding mean square errors of the scalar spatial gradient are given as the ratio of Eq. \ref{eq_wiener_mean_square_error_deriv} and the integral over all wavenumbers of the model dissipation spectrum, $\int_{-\infty}^{+\infty}\!\kappa^{2}S(\kappa)\,\mathrm{d}\kappa$. From Eq. \ref{eq_wiener_mean_square_error}, the mean square error can be thought of as an excess variance contributing to the measurement of the true variance. Similarly, the mean square error of the derivative, Eq. \ref{eq_wiener_mean_square_error_deriv_def}, is the corresponding excess variance to the variance of the spatial derivative, \textit{i.e.} the scalar dissipation. In summary, table \ref{table_wiener_errors} shows that average error of the mixture fraction variance over all times ASI is less than 2\% and that the scalar dissipation rate is accurate to around 20\%. This level of accuracy can be compared to values reported in the literature. For example, in a confined jet experiment the budget of the scalar variance gave an uncertainty within 20\% \cite{mastorakos_sdr_2005} and in a partially stirred reactor \cite{DanailaDimotakis} the accuracy was again within 20\%, judged by the scatter of the values by using different methods to calculate the dissipation rate. The possibility to estimate explicitly the accuracy, Eq. \ref{eq_wiener_mean_square_error_deriv}, and the sensitivity of the filtering process, Eq. \ref{eq_wiener_mean_square_error_suboptimal} below, without relying in ad-hoc assumptions render the present technique suitable for post processing scalar measurements with the aim of calculating the scalar dissipation rate.

\begin{table*}[t!]
\centering
\caption{Percentage (\%) errors of the variances of the mixture fraction and of the scalar spatial derivative, after using the Wiener filter. The errors are normalised with the model scalar variance and the model scalar dissipation, respectively.}
\label{table_wiener_errors}
\begin{tabular}{c|cccccccccc}
Time ASI, t/T             & 0.9  & 2.15 & 3.4  & 4.65 & 5.9  & 7.15 & 9.65 & 10.9 & 13.4 & 15.9 \\
\hline
$\langle\zeta^{2}\rangle$ & 3.5  & 1.9 & 1.6   & 2.4  & 2.3  & 0.8  & 2.0  & 1.0  & 1.5  & 1.3 \\
$\langle\left(\nabla\zeta
\right)^{2}\rangle$       & 19.2 & 16.1 & 29.2 & 18.2 & 23.3 & 10.7 & 19.9 & 13.9 & 22.6 & 23.6\\
\end{tabular}
\end{table*}

Finally, we comment on the effect of the spectral fittings that are used to define models for the noise spectrum and the high wavenumber signal spectrum and show that suboptimal models affect little the derived quantities. Suppose that the Wiener filter was estimated with model spectra that were different from the real spectra, thus producing a filter function $W_{*}$ instead of the optimum one, $W$. Then, the mean square error ($e^{2}_{*}$) for this choice of Wiener filter is given by Eq. \ref{eq_wiener_mean_square_error_expand}, where $W$ is replaced by $W_{*}$. Expanding the expression inside the braces and then completing the square in the ensuing expression results in
\[
e^{2}_{*}=\int_{-\infty}^{+\infty}\!
\left[
\frac{1}{H^{2}}\left(F_{S}^{2}+N^{2}\right)
\left(W_{*}-W\right)^{2} +
\frac{1}{H^{2}}\frac{F_{S}^{2}N^{2}}{F_{S}^{2}+N^{2}}
\right]\,\mathrm{d}\kappa
\]
The last term inside the braces in the above equation is the mean square error for the optimum Wiener filter, Eq. \ref{eq_wiener_error}. So the mean square error for the "sub-optimal" Wiener filter is given by
\[
e^{2}_{*}=e^{2} + \int_{-\infty}^{+\infty}\!\frac{F_{S}^{2}+N^{2}}{H^{2}}
\left(W_{*}-W\right)^{2}\,\mathrm{d}\kappa
\]
This expression is quadratic in the difference between the two Wiener filters, implying that even a non-perfect evaluation of the model spectra is still capable of producing small filtering errors, comparable with the minimum one. For example, $W_{*}=(1+\alpha)W$, where $\alpha$ is assumed constant for all wavenumbers. From Eq. \ref{eq_wiener_filtered_image} it follows that $W^{2}/H^{2}=F_{F}^{2}/F_{M}^{2}=F_{F}^{2}/\left(F_{S}^{2}+N^{2}\right)$ and the above equation becomes
\begin{equation}
e^{2}_{*}=e^{2} + \alpha^{2}\int_{-\infty}^{+\infty}\!F_{F}^{2}\,\mathrm{d}f
\label{eq_wiener_mean_square_error_suboptimal}
\end{equation}
so the extra mean square error for, say, $\alpha=0.1$ (or 10\%) is 1\% of the measured variance for both the mixture fraction and the scalar spatial derivative. Therefore, the current analysis shows that the Wiener filter is rather insensitive in the choice of models, a fact that was verified for some cases by choosing alternative models for either the measurement or the noise spectra.

In summary, the Wiener filter tries to recreate the noise-contaminated turbulent signal by forcing an agreement (in the least mean square sense) between the measured and the "ideal" spectra. In doing so, the scale-dependent effect of noise is taken into account in order to reconstruct the noise-free signal. It is still important, however, to analyse the effectiveness of the process and this is done in section \ref{section_results}.

\subsection{A dissipation correction method based on Richardson's extrapolation} \label{section_eaton_correction_method}
The method of \cite{eaton_2007_dissipation_correction_piv} provides a correction to energy dissipation rate measurements (easily extendible to scalar dissipation rate as well) for the effects of noise and inadequate spatial resolution. Since it can be shown that this method is second order accurate in the spatial resolution (thus giving an estimate of the accuracy), we briefly present it here and use it in subsequent sections to compare it with the Wiener filter, in trying to render greater credibility to the modelling assumptions of the Wiener filter adopted earlier. This method, however, can only correct for the mean dissipation rate, so it cannot be used when the instantaneous fields are required.

The idea of Richardson extrapolation is that by repeating a numerical calculation with different values of a parameter (in this case the spatial resolution), the 'true' value of the quantity that needs to be calculated can be obtained by extrapolation to zero value of the parameter. In \cite{eaton_2007_dissipation_correction_piv} this approach is used to calculate the turbulent kinetic energy dissipation rate from Particle Image Velocimetry (PIV) images, where the spatial resolution is adequate and experimental noise dominates. The energy dissipation rate is evaluated at two spatial resolutions, $\Delta x$ and $2\Delta x$, and a combination of these two measurements gives a measurement of the ensemble averaged dissipation rate which is second order accurate on $\Delta x$. It is straightforward to adapt this correction method to the scalar dissipation rate and the formula to calculate the mean scalar dissipation rate is
\begin{equation}
\left<\chi\right>=\frac{4\left<\chi_{2\Delta}\right> - \left<\chi_{\Delta}\right>}{3}
\label{eq_dissipation_correction_eaton}
\end{equation}
where $\chi_{2\Delta}=\left(\Delta\zeta/\Delta x\right)_{2\Delta x}^{2}+\left(\Delta\zeta/\Delta y\right)_{2\Delta y}^{2}$ and the terms are calculated as, for example,
\[
\left(\frac{\Delta\zeta}{\Delta x}\right)_{2\Delta x}=\frac{\zeta_{i+2,j}-\zeta_{i-2,j}}{4\Delta x}
\]
where $i,j$ denote the pixels and $\Delta x,\Delta y$ the pixel spacings in the two directions.

%%%%%%%%%%%%%%%%%
\section{Results} \label{section_results}
In this section the effects of filtering on the measured mixture fraction distributions are presented and results from the starting jet flow are reported in order to demonstrate the measurement approach.

\subsection{Filtering model noise} \label{section_wiener_method_noise}
The portion of the spectrum that is modelled through the exponential decay in the construction of the Wiener filter should be as small as possible. Furthermore, the properties of the residual fields (measured minus filtered) have to be tested to check the effectiveness of the filtering operation.

The level of modelling of the measured spectrum (figure \ref{fig_wiener_model_spectra}), when determining the Wiener filter, can be given by the portion of the measured variance that is modelled by the large-wavenumber exponential decay. Using the same procedure as earlier (\textit{c.f.} figure \ref{fig_noise_spectrum_explain}) this portion is given by the ratio of the area between the measured and fitted spectra over the area under the measured spectrum. The 'cut-off' wavenumber, where the modelled spectrum starts to depart from the measured spectrum is of the same order as the smallest estimated Batchelor length scale, which is true for other times ASI as well, showing that the modelling does not affect the smallest scales of the scalar field.
%The same procedure is followed for the 'dissipation' spectrum (calculated as $D(\kappa)=\kappa^2E(\kappa)$, where $\kappa=\sqrt{u^2+v^2}$ is the wavenumber and $E(\kappa)$ is the scalar energy spectrum).
Table \ref{table_wiener_modeling_percentage} shows the relevant percentages of modelled variance and the 'cut-off' wavenumbers, for all times ASI. The mixture fraction variance is almost entirely determined by the measurements, with the fitted part of the spectrum being less than 1\% in all cases. This should be expected given the relatively high SNR calculated earlier. The large 'cut-off' length scales for the first couple of times ASI are a consequence of the development of the scalar field at these earlier times, where the fluid has just left the nozzle and the scalar field has not had time to develop turbulent fluctuations. The spectrum is determined only by the very large scales, with no intermediate scales present, so the noise contribution extends at relatively low wavenumbers.

\begin{table*}[t!]
\centering
\caption{The cut-off wavenumber (\unit[]{mm$^{-1}$}), where the model spectrum starts to deviate from the measured spectrum and percentage (\%) of the measured spectrum of the mixture fraction that is modelled in the Wiener filter.}
\label{table_wiener_modeling_percentage}
\begin{tabular}{c|cccccccccc}
Time ASI, t/T                       & 0.9  & 2.15 & 3.4  & 4.65 & 5.9  & 7.15 & 9.65 & 10.9 & 13.4 & 15.9 \\
\hline
$\kappa_{\mathrm{cut-off}}$         & 1.2  & 1.9  & 3.5  & 2.5  & 3.0  & 2.7  & 2.9  & 2.8  & 1.7  & 2.6 \\
$E(\kappa)_{\mathrm{modelled}}$     & 0.26 & 0.08 & 0.01 & 0.03 & 0.02 & 0.01 & 0.01 & 0.01 & 0.03 & 0.02
%$D(\kappa)_{\mathrm{modelled}}$     & 49.8 & 18.1 & 4.1  & 7.3  & 6.4  & 4.2  & 5.3  & 7.4  & 52.0 & 20.4\\
\end{tabular}
\end{table*}

\begin{figure*}[htbp]
\centering
\begin{tabular}{ccc}
\includegraphics[scale=0.4]{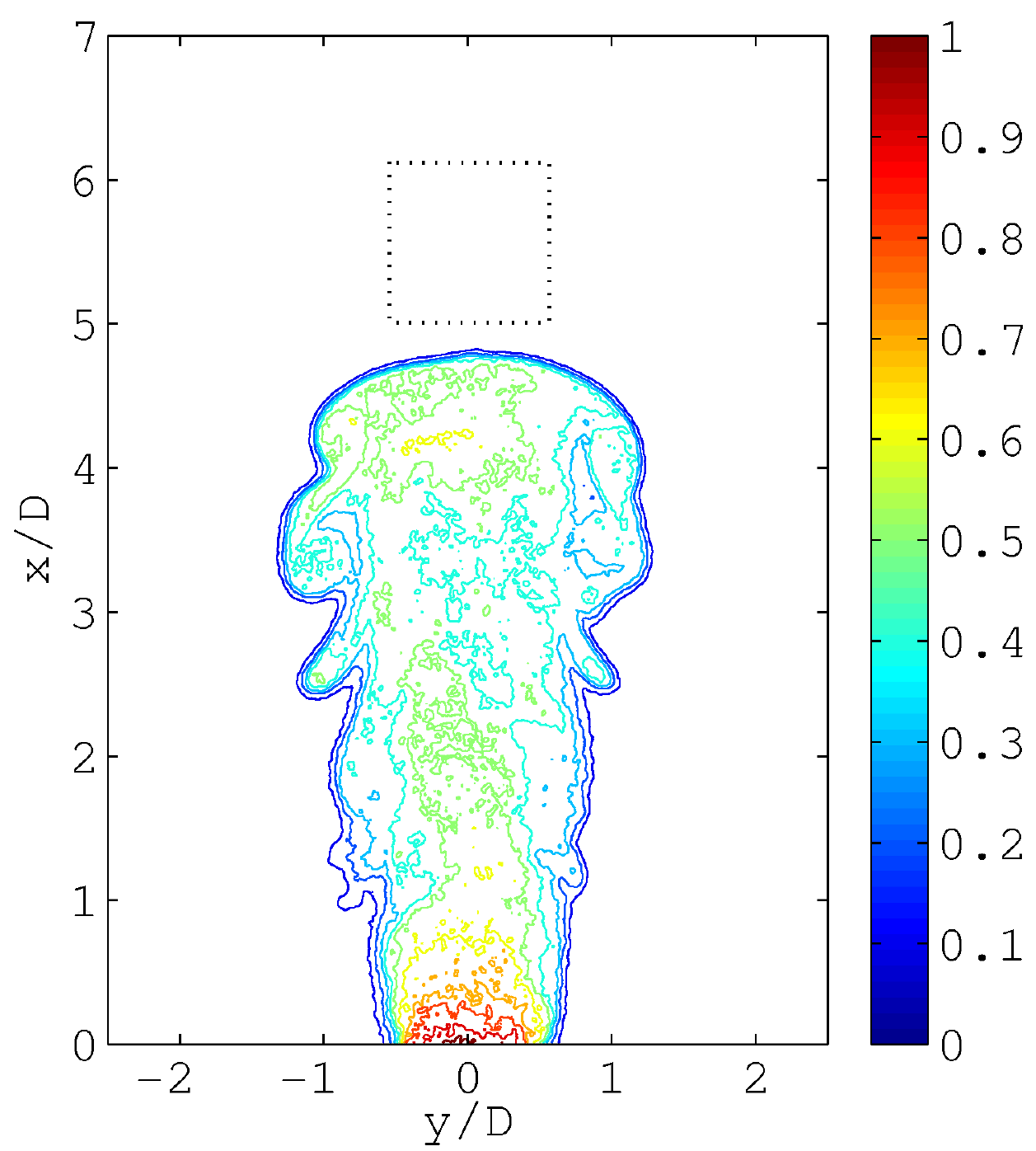} &
\includegraphics[scale=0.4]{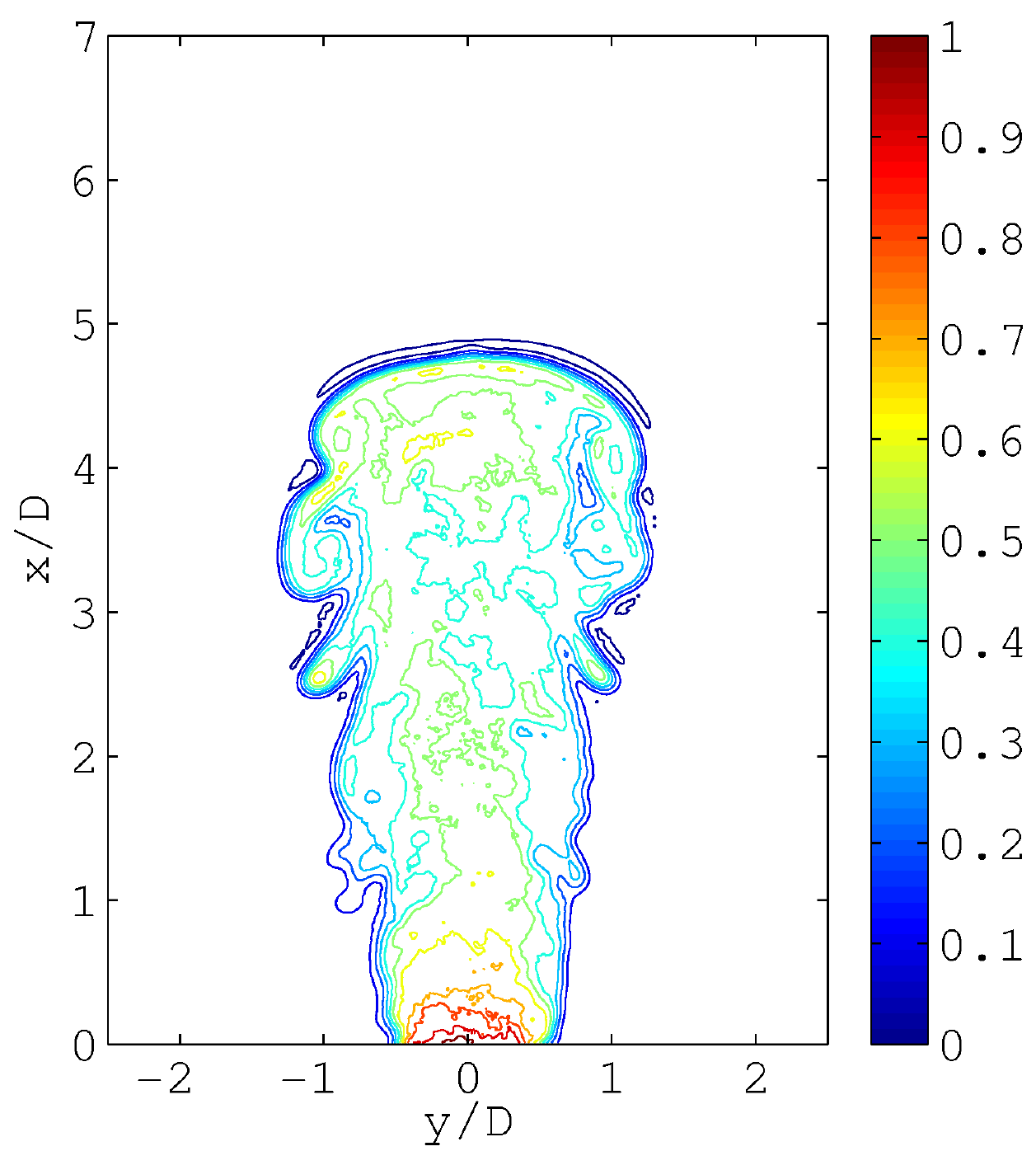} &
\includegraphics[scale=0.4]{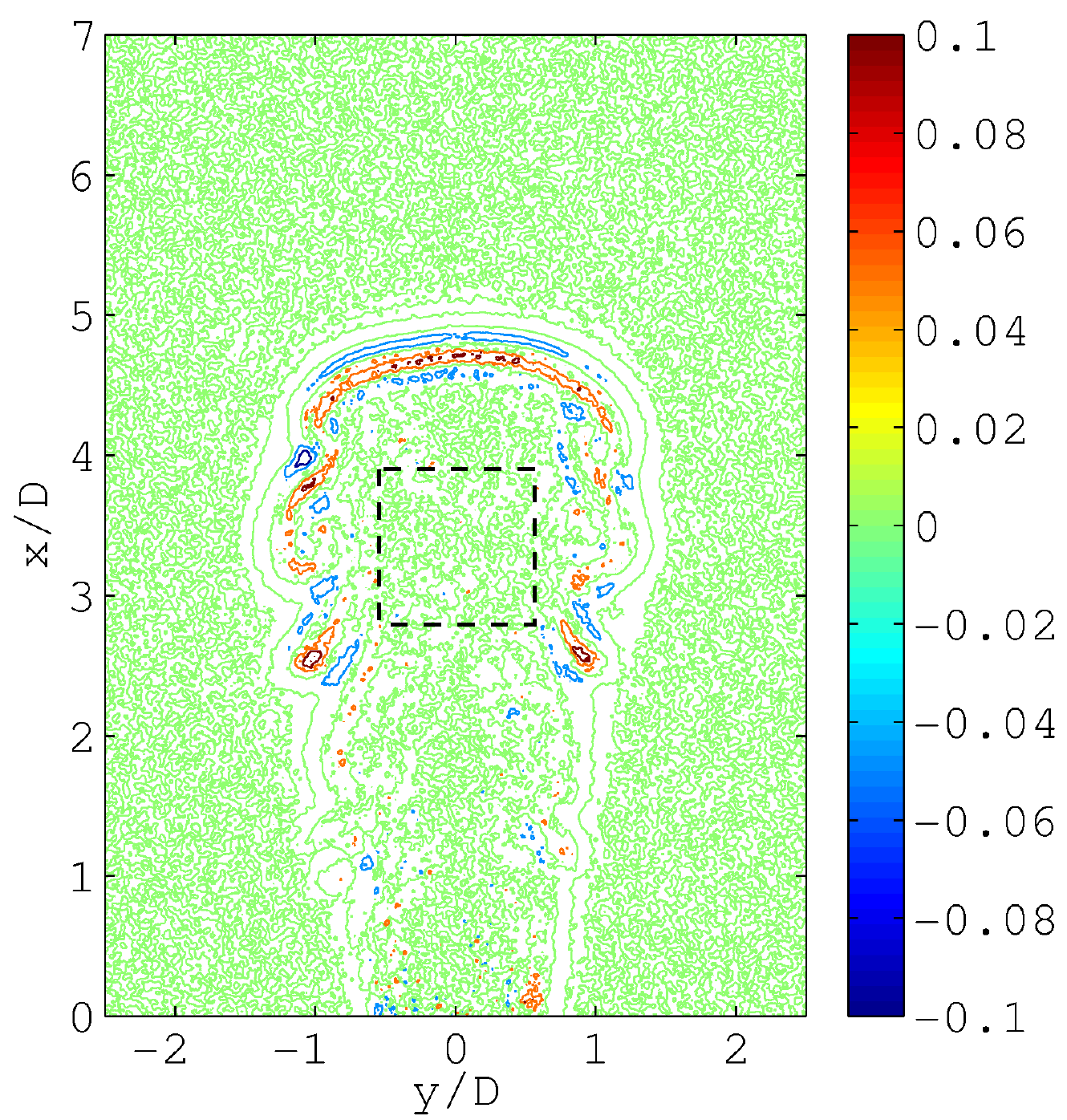}
\end{tabular}
\caption{An instantaneous raw mixture fraction image at time t/T=9.65 (left), the corresponding instantaneous image after application of the Wiener filter (centre) and the difference between the instantaneous images, filtered minus raw (right). The scale corresponds to mixture fraction.}
\label{fig_images_raw_wiener_diff}
\end{figure*}

In assessing the statistical properties of the corrections made to the raw images, figure \ref{fig_images_raw_wiener_diff} presents a raw instantaneous mixture fraction image at time t/T=9.65, the same instantaneous image after the application of the Wiener filter and the difference between these two images (raw minus filtered).

The two instantaneous images are practically the same to each other (as shown later the correction is less than 2\% of the measured mixture fraction values and, typically, less than 1\%) and the general characteristics of the evolving jet are not changed by the application of the filter. In the filtered image the contour lines are smoother without altering their location and shape. Qualitatively, this shows that the Wiener filter is adequate in removing noise contributions, with more quantitative results given later in this and the next subsections.

\begin{figure}[htbp]
\centering
\includegraphics[scale=0.4]{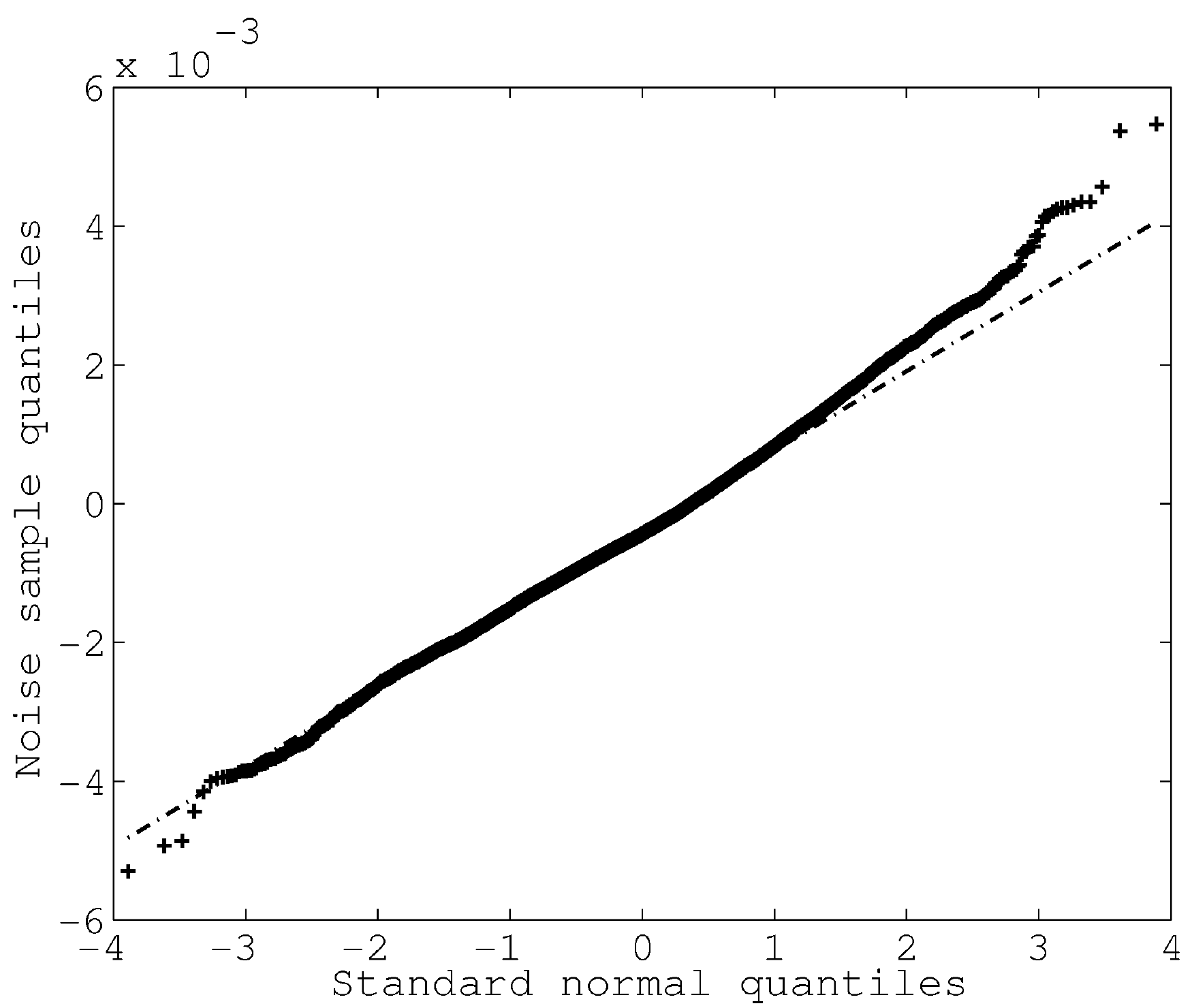}
\caption{Quantile-quantile plot for the camera noise. The noise sample is produced from an image of the deviations from the ensemble average for a region outside the jet flow.}
\label{fig_noise_qqplot}
\end{figure}

Initially, the noise characteristics of the imaging system are assessed from the recorded images. The viewing area in all images contains the jet plus regions of the laser sheet where no acetone vapour is present; in these regions the acquired signal should be constant with only the camera noise superimposed. For example, subtracting the ensemble averaged image from a raw instantaneous image and collecting the pixel values inside the rectangle shown in the leftmost plot of figure \ref{fig_images_raw_wiener_diff}, will give the noise distribution of the camera. Figure \ref{fig_noise_qqplot} shows the distribution of the noise samples from this region compared with a standard normal distribution (the quantile plot is constructed by plotting ordered data values of two distributions against each other; in case the two data sets follow the same distribution the plot is a straight line). For most part of the graph the noise sample appears to be close to the normal distribution, at least up to 2 standard deviations and noise samples from different instantaneous images and for different times ASI show a similar behaviour. So, we can assume that the noise characteristics of the camera follow a gaussian distribution.

The image obtained as a result of the filtering operation is the best estimate of the unblurred, noise-free image. So, the noise field (the difference between the filtered and the raw images -- \textit{c.f.} the image model, Eq.\ref{eq_image_model}) should be both as small as possible and as close to white noise as possible. It should be arguably accepted that the level of the correction should be as small as possible and in line with the signal-to-noise ratio. Also, the filtering should preserve the features of the measured flow field and not remove any large or smaller scale characteristics. So, it should be a featureless noise field, \textit{i.e.} white noise. The difference image in figure \ref{fig_images_raw_wiener_diff} shows some remaining 'structure' after filtering. The level of this residual structure is small, less than 3\% of the measured value, however it points, as explained later, to one of the limitations of the filtering technique.

\begin{figure}[htbp]
\centering
\includegraphics[scale=0.4]{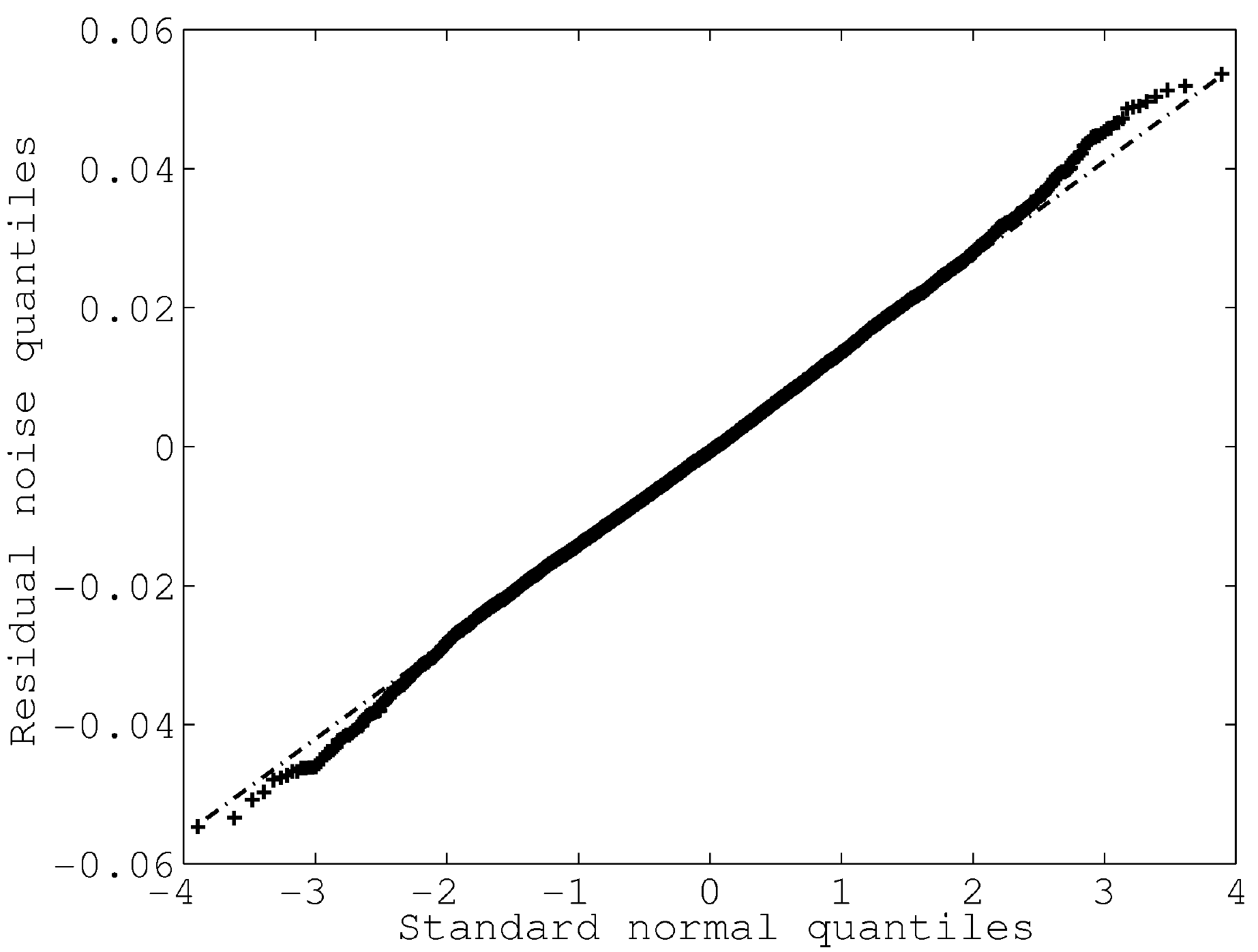}
\caption{Quantile-quantile plots of the optimal Wiener filter model noise compared with a standard normal distribution.}
\label{fig_noise_filters_norm_qqplot2}
\end{figure}

A noise distribution is shown in figure \ref{fig_noise_filters_norm_qqplot2}. The sample is obtained by collecting all the pixels inside the small rectangle in the rightmost image of figure \ref{fig_images_raw_wiener_diff}, the difference between filtered and measured images. The noise distribution compares well with the normal distribution with only minimal deviations from the gaussian distribution at fluctuations larger than about 2 standard deviations. The distributions at other positions in the flow (from regions within the jet contour) and at other times ASI are very similar. This particular noise distribution has a mean value of zero and variance $10^{-4}$; in general, the level of the correction is less than 1\% of the measured value of the mixture fraction. This level can be assessed by considering, for example, that when the noise is gaussian and additive, at every spatial location in the image its distribution is $N(0,\sigma^{2}/SNR)$, where $N(a,b)$ is the normal distribution with mean $a$ and variance $b$ and $\sigma$ is the standard deviation of the measurement at the same spatial location. So, for a typical value of the standard deviation of the mixture fraction fluctuations $\sigma\approx0.1$ and $SNR\approx100$ the variance of the noise should be $\approx10^{-4}$, as measured. Consequently, the correction imposed by the Wiener filter is generally small and consistent with the signal-to-noise ratio of the measurements and the distribution of the residual noise samples follows well a normal distribution. This means that the Wiener filter does not alter the underlying mixture fraction field and, more importantly, does not introduce any bias in the values of the corrected images. This figure also verifies the assumption that the camera noise follows a gaussian distribution.

\begin{figure}[htbp]
\centering
\includegraphics[scale=0.45]{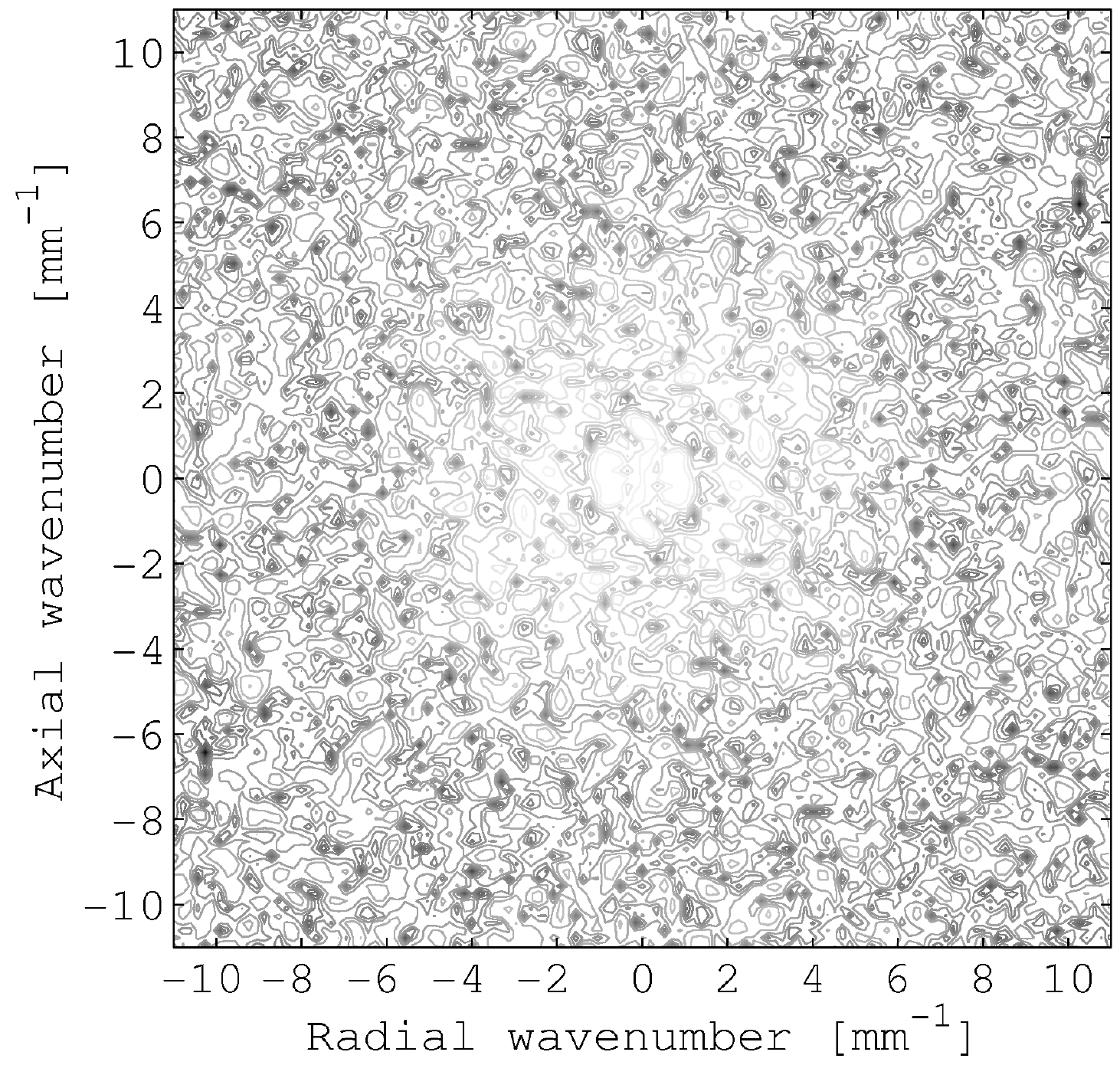}
\caption{The 2D spectrum of the residual noise inside the rectangle in the rightmost image of figure \ref{fig_images_raw_wiener_diff}. The logarithm base-10 of the spectrum is shown in this figure.}
\label{fig_noise_residual_2d_spectrum}
\end{figure}

The 2D spectrum of the noise values in the same small rectangle of the rightmost image of figure \ref{fig_images_raw_wiener_diff}, is shown in figure \ref{fig_noise_residual_2d_spectrum}. This spectrum has been calculated from a single image, which explains the random fluctuations evident in this plot and the ensemble averaged spectrum shows the same characteristics as the single-shot spectrum. Small regions at other positions inside the jet body and for other times ASI give similar results. This spectrum is symmetric around the origin and at the larger wavenumbers the magnitude of the spectrum is almost the same everywhere, irrespective of orientation or scale. This implies that there is no bias in the filtering of the very small scales. At larger scales, larger than about \unit[0.4-0.5]{mm}, there is a small increase in the level of the spectrum. This scale corresponds roughly to the cut-off wavenumber where the model spectrum starts to deviate from the measured spectrum of the mixture fraction fluctuations, table \ref{table_wiener_modeling_percentage}. At wavenumbers smaller than the cut-off wavenumber, the noise level is considerably smaller than the signal. However, the filtering operation still attenuates the measured signal. In this way, some signal structure is smoothed out, as seen in figure \ref{fig_images_raw_wiener_diff}. In contrast, at larger wavenumbers where noise starts to become comparable to the measured signal the Wiener filter acts to reconstruct the ideal signal by only removing the noise contribution.

Finally, we note that the above discussion referred to flow regions away from the jet boundaries. Outside and inside the jet body the residual noise distribution is featureless and the ensuing corrections are small (as shown by the noise distribution in figure \ref{fig_noise_filters_norm_qqplot2} and the 2D spectrum in figure \ref{fig_noise_residual_2d_spectrum}). Despite that, the rightmost plot of figure \ref{fig_images_raw_wiener_diff} shows that there is some dampening of characteristic features of the mixture fraction field at the jet boundaries. Since the jet boundaries always have very large gradients, these gradients will always be attenuated after applying the filter. So, given the persistence of these gradient from injection to injection (in contrast to the random emergence of large gradients inside the jet body), the scalar dissipation rate will have a negative bias at the jet boundaries.

\subsection{Scalar dissipation rate measurements}
This section presents results of the scalar dissipation rate measurements, compares the raw with the filtered images and proves the effectiveness of the Wiener filter in obtaining accurate scalar dissipation rate measurements.

\begin{figure*}[htbp]
\centering
\begin{tabular}{cc}
\includegraphics[scale=0.4]{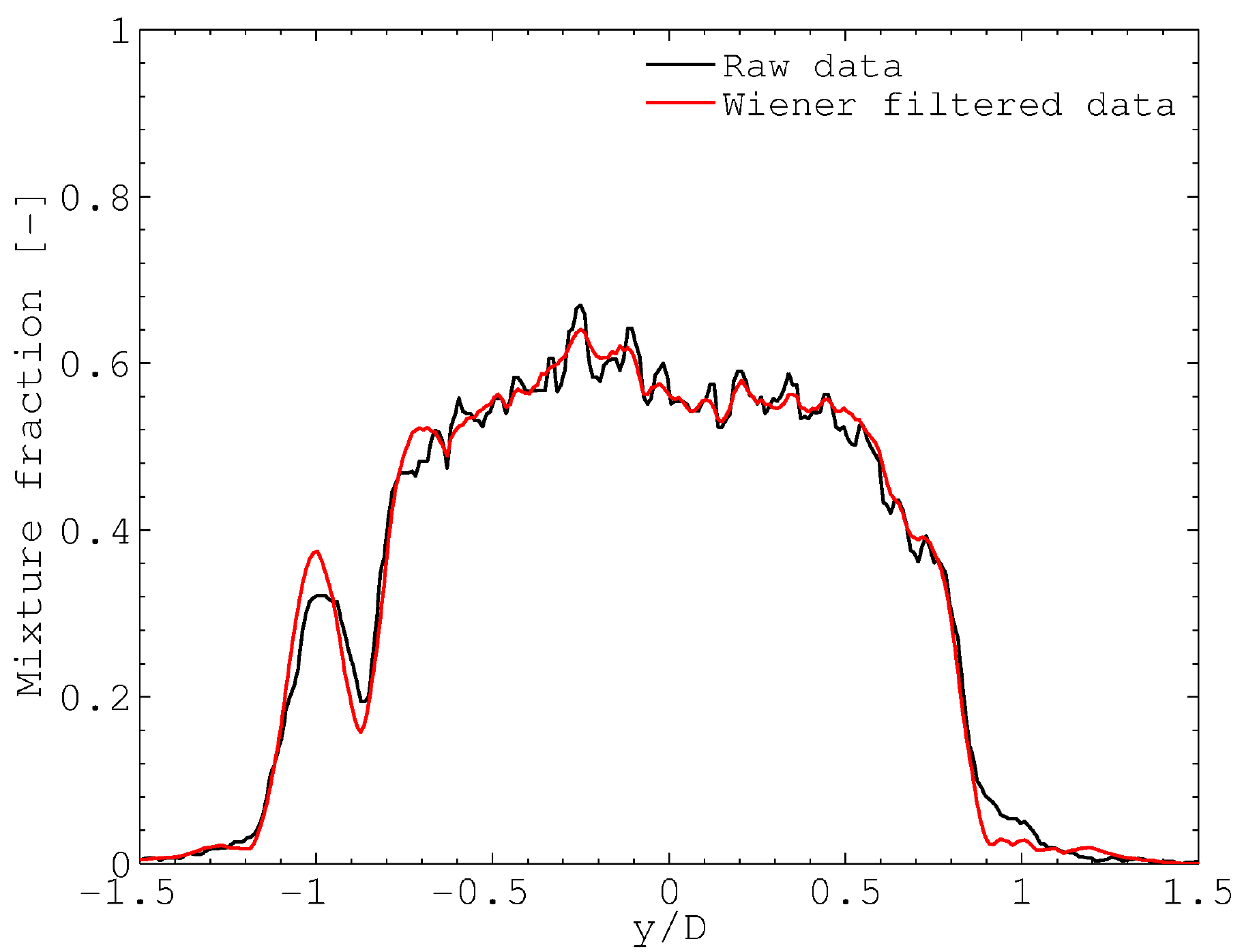} &
\includegraphics[scale=0.4]{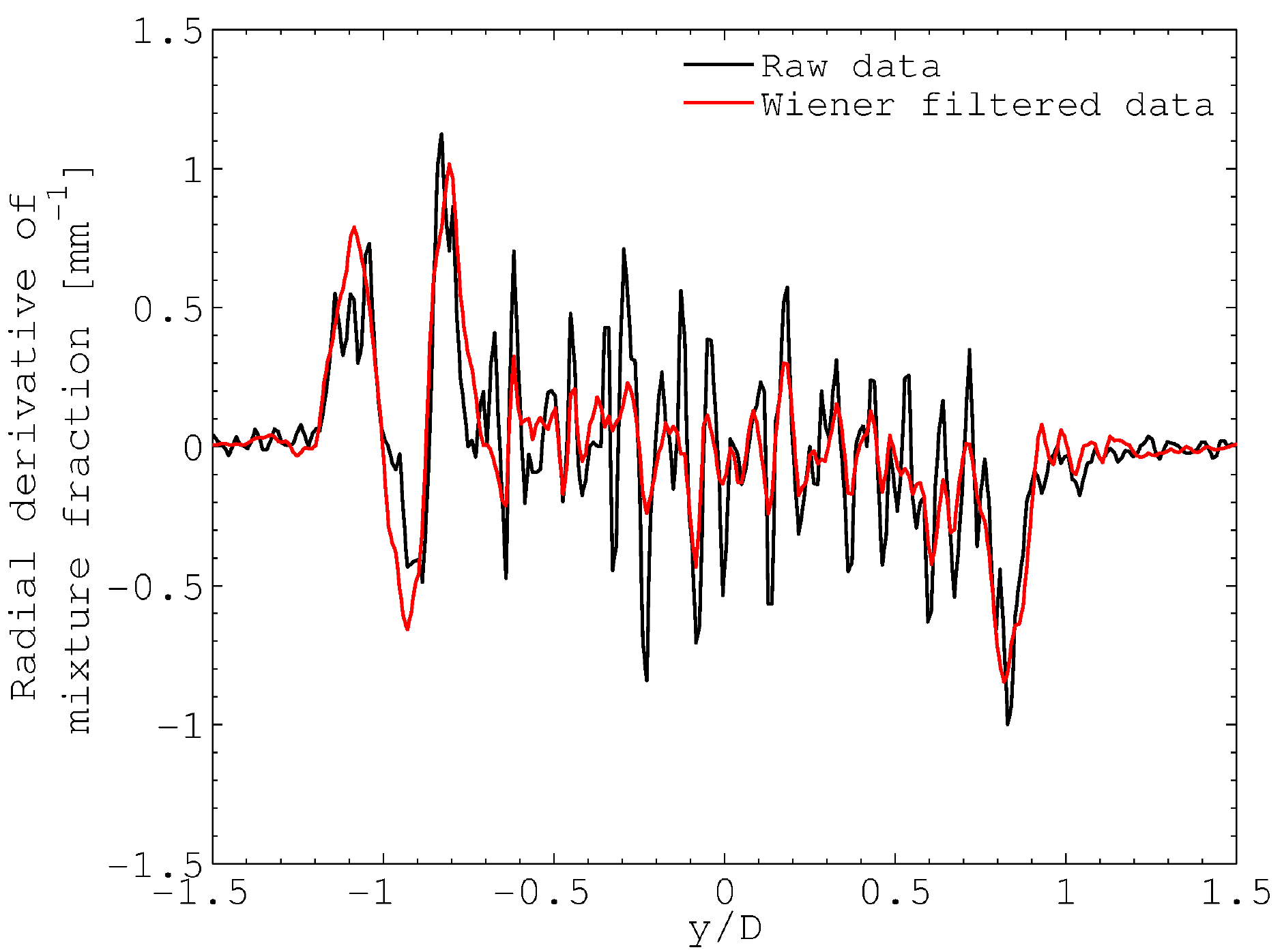}
\end{tabular}
\caption{Radial profiles of an instantaneous raw mixture fraction image at time t/T=10.9 and x/D=4 and of the corresponding filtered image (left). Corresponding radial derivatives of the same profiles (right).}
\label{fig_wiener_filter_profiles_examples}
\end{figure*}

The scalar dissipation rate will be presented without the diffusivity, so its dimensions are in \unit[]{mm$^{-2}$}. We can assume that the diffusivity is constant throughout the flow with a value between \unit[20]{mm$^2$/s} and \unit[25]{mm$^2$/s}. In relation to figure \ref{fig_images_raw_wiener_diff}, figure \ref{fig_wiener_filter_profiles_examples} shows radial profiles, at the same axial distance, of both the raw and filtered image and, also, of the radial derivative of both signals. The Wiener filter proves effective in restoring the signal and removing small fluctuations that could be attributed to noise.

\begin{figure*}[htbp]
\centering
\begin{tabular}{cc}
\includegraphics[scale=0.45]{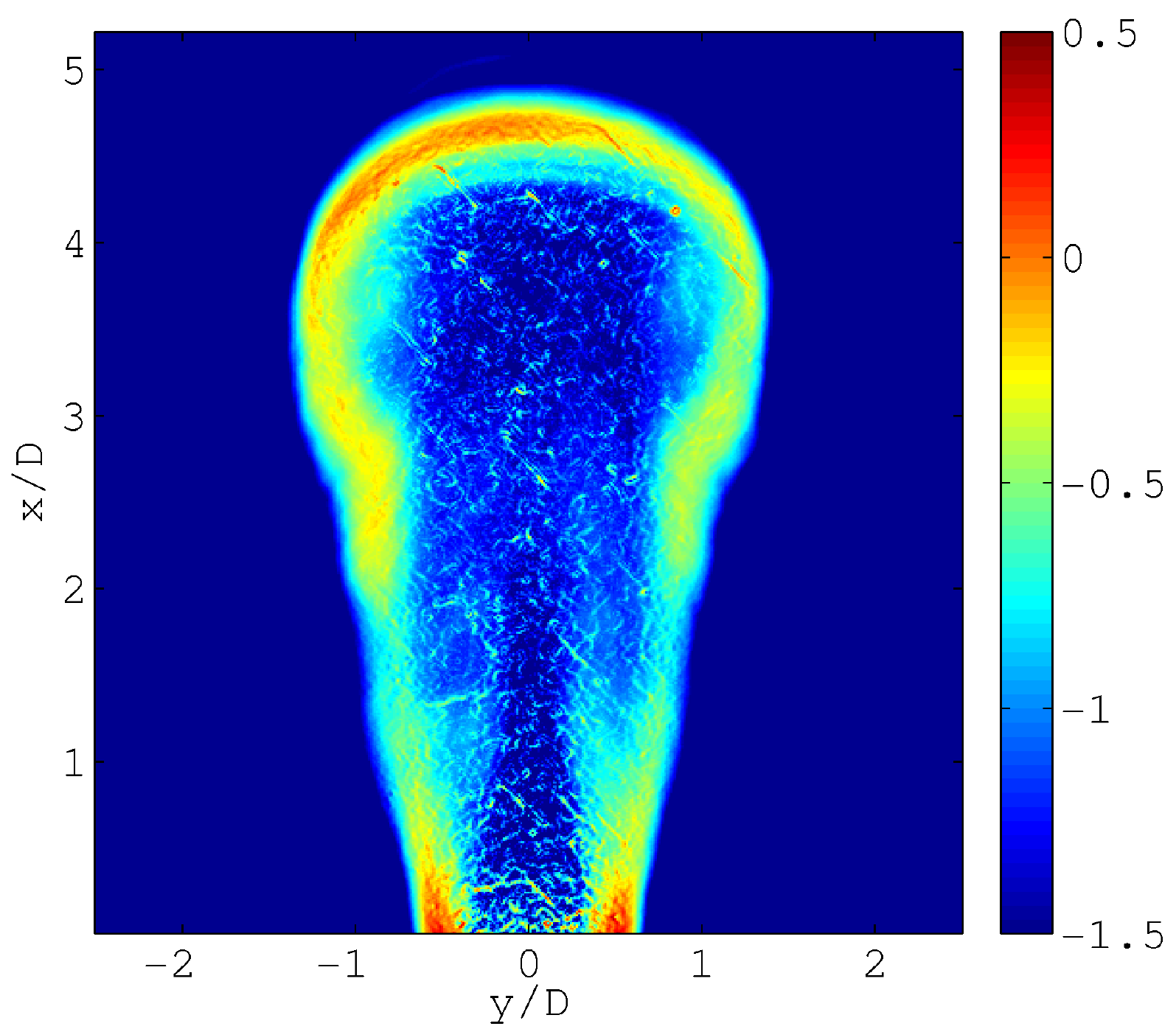} &
\includegraphics[scale=0.45]{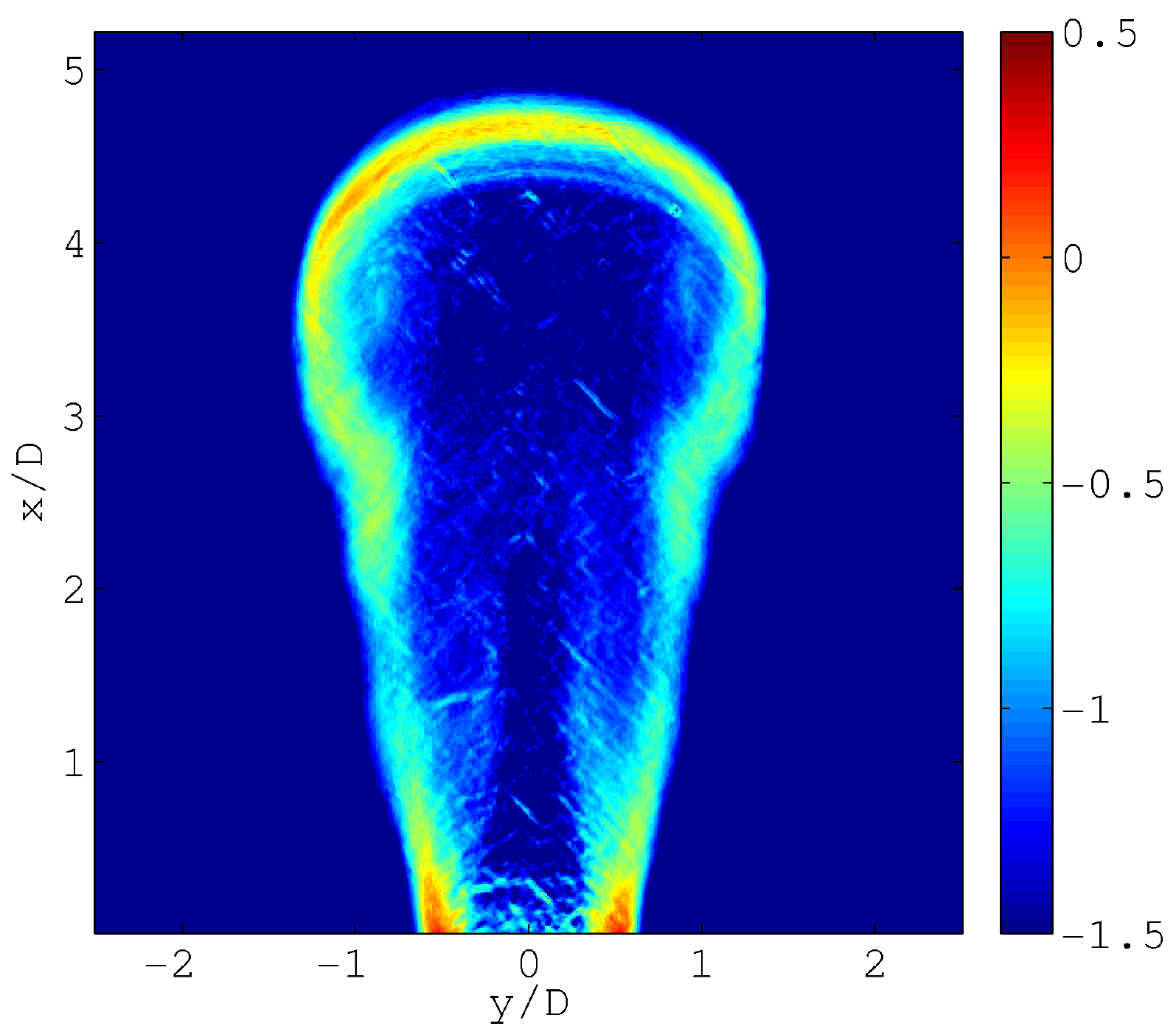}
\end{tabular}
\caption{The mean scalar dissipation rate at time t/T=9.65 for the Wiener filter (left) and the correction method from \cite{eaton_2007_dissipation_correction_piv} (right). The images show the logarithm base-10 of the scalar dissipation without the diffusivity.}
\label{fig_mean_dissipation_wiener_eaton}
\end{figure*}

In order to assess the ability of the Wiener filter to produce accurate results for the mean scalar dissipation rate, figure \ref{fig_mean_dissipation_wiener_eaton} shows the mean scalar dissipation rate at time t/T=9.65 for the Wiener filter and the correction method from \cite{eaton_2007_dissipation_correction_piv}, as presented in section \ref{section_eaton_correction_method}. The two distributions are very similar to each other with differences being found in some structures within the jet body, which have very small dissipation values; it is, also, important to note that the values of the mean scalar dissipation rate, as calculated from the two methods, are almost the same. The Wiener filter practically, and as applied here, corrects for the noise in the images, whereas the correction method from \cite{eaton_2007_dissipation_correction_piv} corrects for both noise and resolution effects. By taking into account figure \ref{fig_mean_dissipation_wiener_eaton}, where the two methods give almost the same results, we can conclude that the spatial resolution of the measurements is enough to resolve the smallest scales of variation of the scalar field with a high signal-to-noise ratio and that we can, consequently, measure the scalar dissipation rate with good accuracy.

\begin{figure*}[htbp]
\centering
\begin{tabular}{cc}
\includegraphics[scale=0.35]{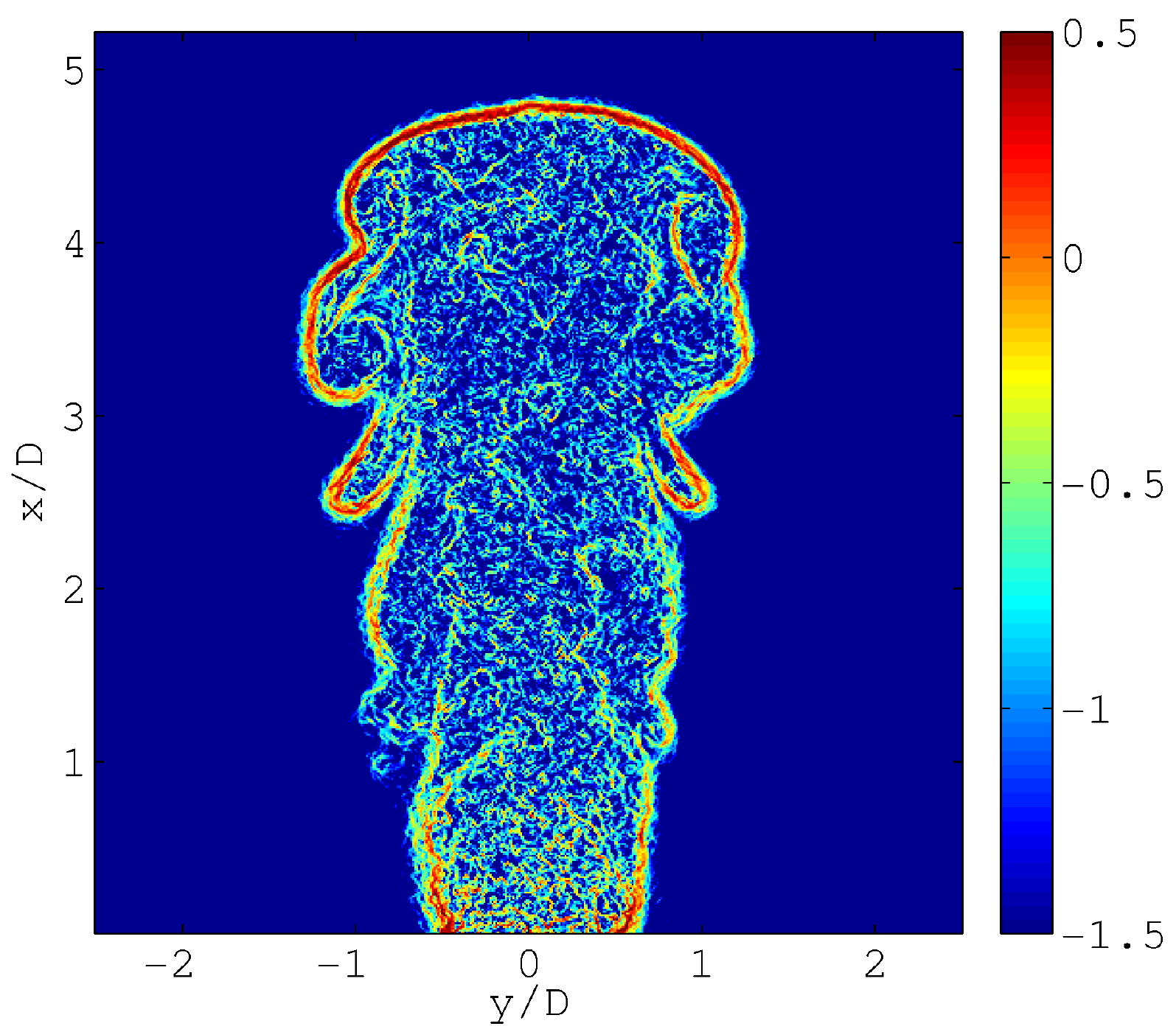} &
\includegraphics[scale=0.35]{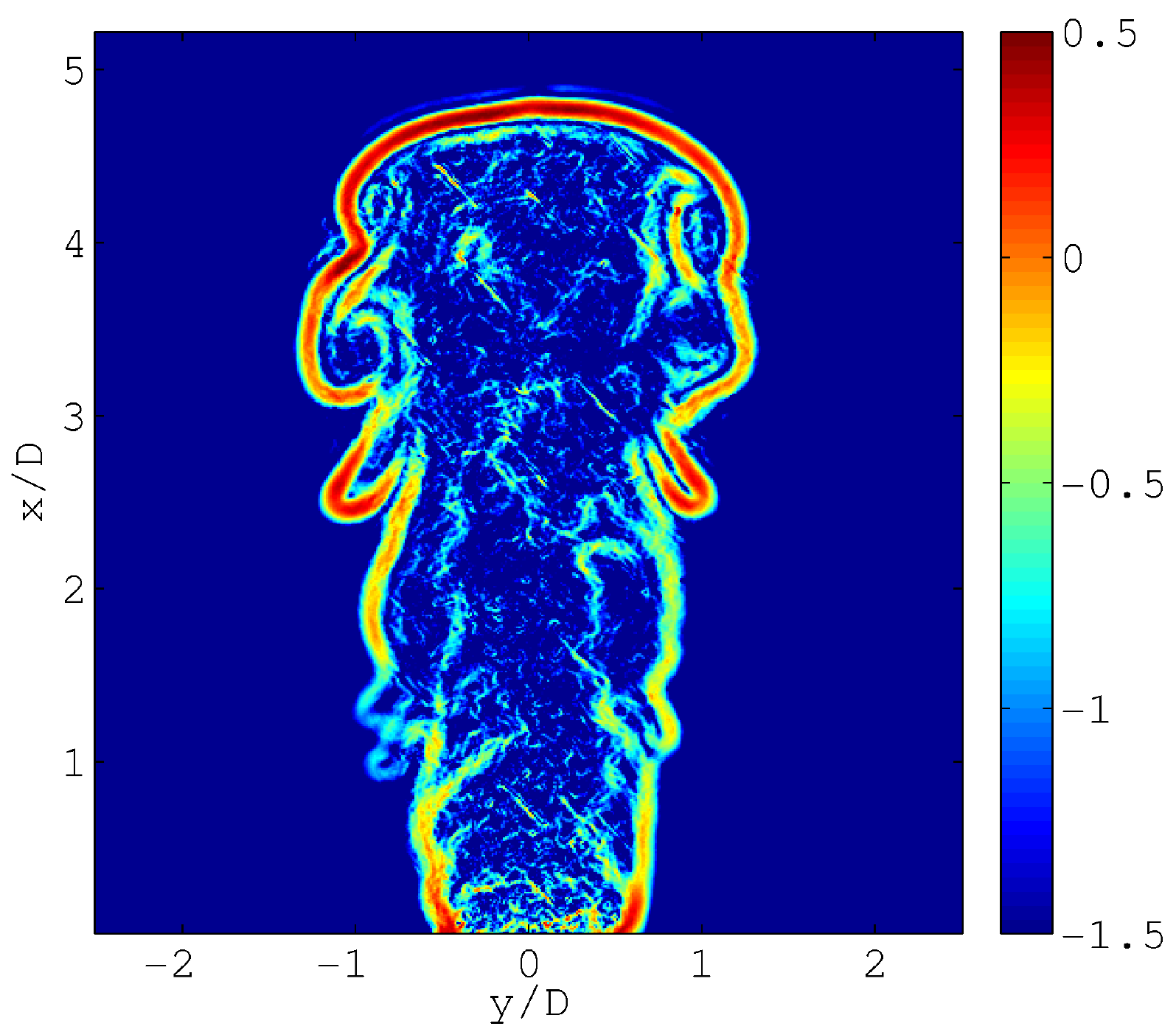} \\
\includegraphics[scale=0.35]{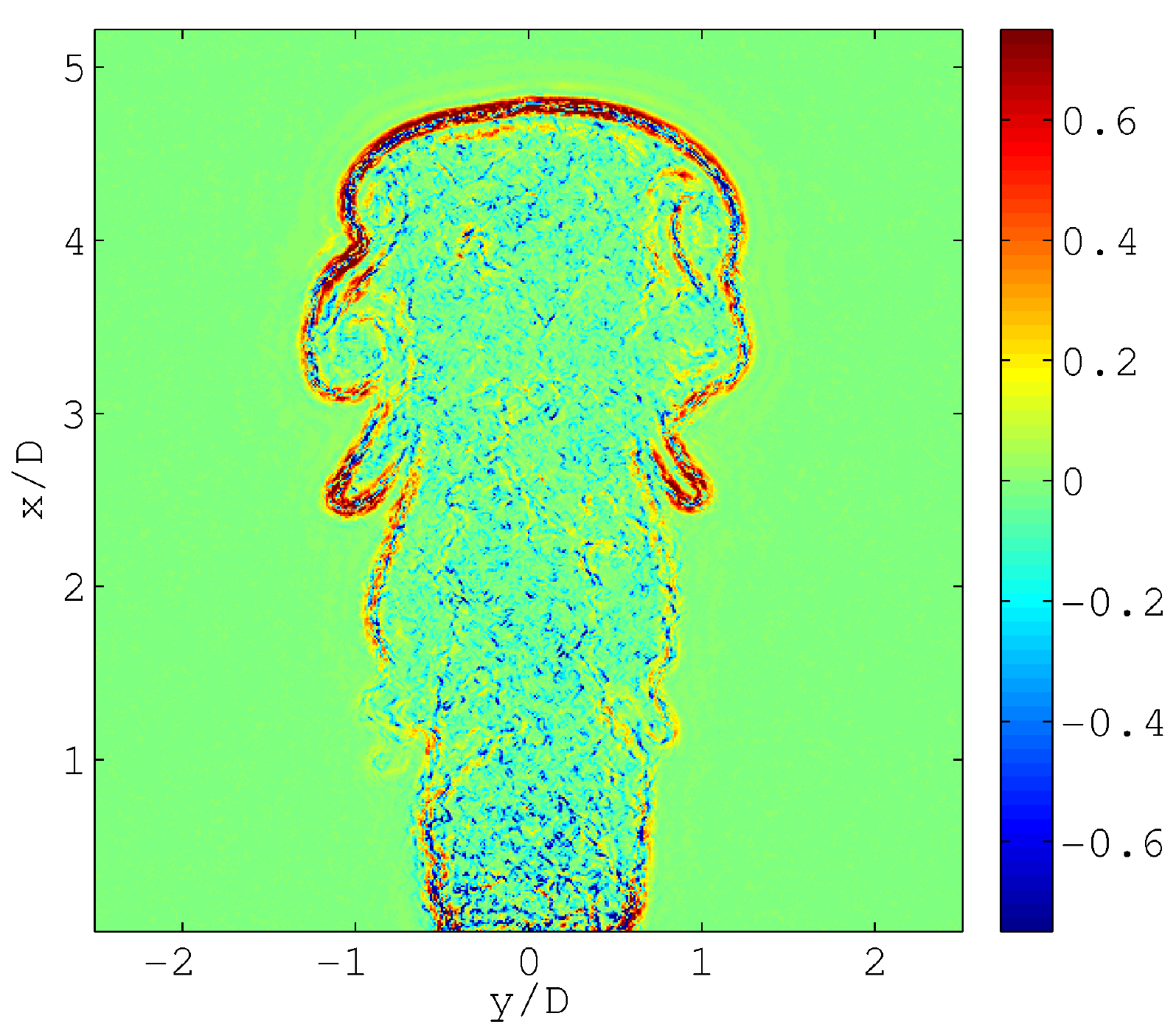} &
\includegraphics[scale=0.35]{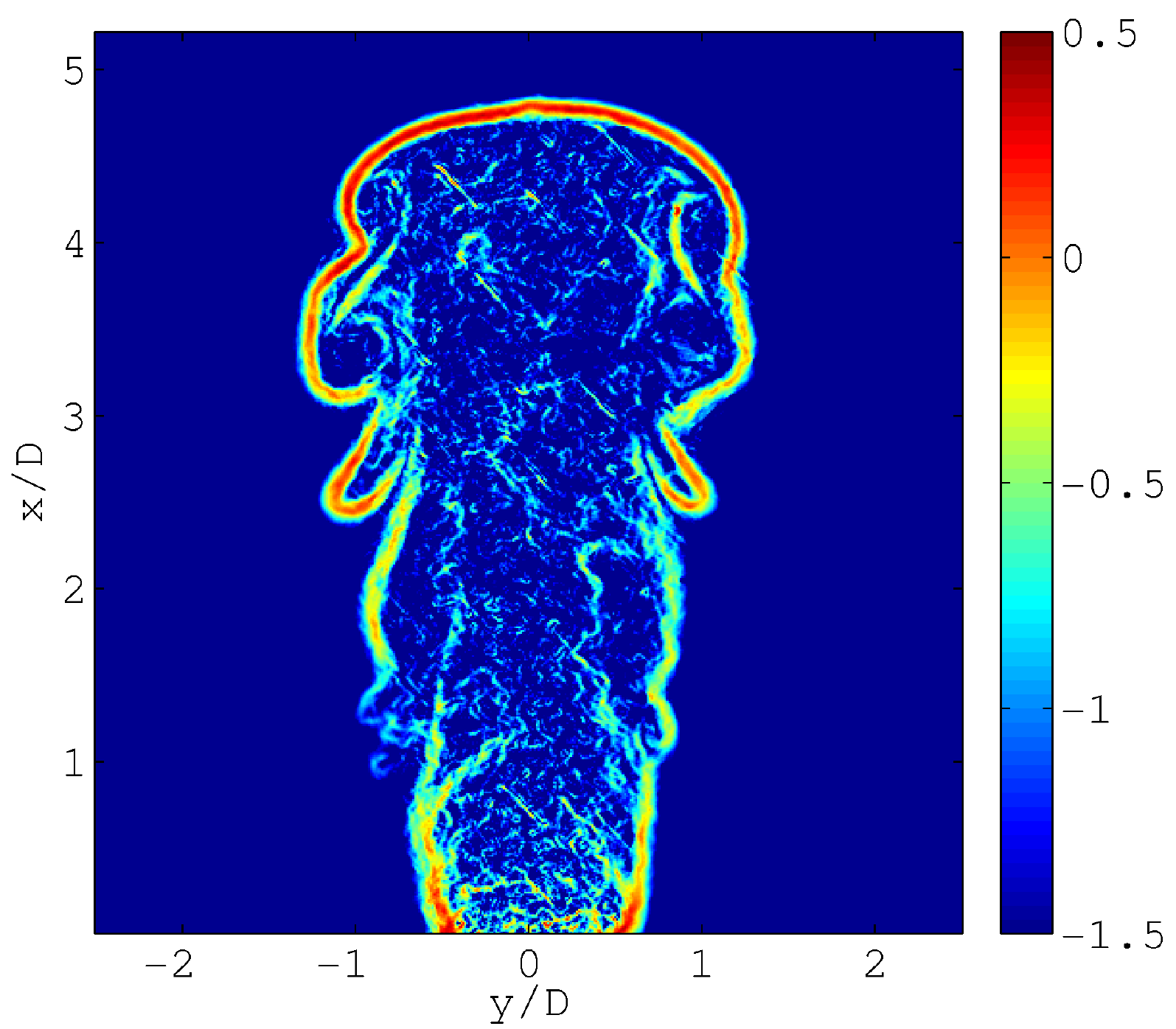}
\end{tabular}
\caption{An instantaneous raw scalar dissipation image at time t/T=9.65 (top left), the corresponding image after application of the Wiener filter (top right) and the difference between the filtered and the raw images (bottom left). Also shown is the instantaneous image assuming the PSF is a delta function (bottom right). The images show the logarithm base-10 of the scalar dissipation without the diffusivity; the difference image shows the difference of the dissipation values without the diffusivity.}
\label{fig_dissipation_raw_wiener_inst}
\end{figure*}

Instantaneous scalar dissipation images of both the raw and the filtered measurements are shown in the top row of figure \ref{fig_dissipation_raw_wiener_inst}. These distributions show a thin layer of very high scalar dissipation values surrounding the jet. Inside the jet body dissipation values are found in small regions within an otherwise calm fluid and the level of the dissipation values are somehow smaller than the values found at the jet boundary. The distribution of the scalar dissipation rate within the jet boundaries changes from injection to injection with the result being a distribution of mean scalar dissipation rate having two peaks at jet shear layer with very low values in between, as shown in figure \ref{fig_mean_dissipation_wiener_eaton}. Part of the thickness of the ensemble average scalar dissipation profile along the radial direction, especially for further axial distances, could result from large scale variations between injections. We will quantify the effect, in subsequent analysis, by considering the repeatability of consecutive injection events through the distributions of the mixture fraction at locations near the nozzle exit, where interaction with the surrounding fluid is minimal; in general, the distributions at these locations are single-mode and compact.

Qualitatively in the raw instantaneous image any structure is always hidden behind a noise 'veil'. In contrast, in the filtered image, this noise has been largely removed and the expected smooth regions of very low dissipation values appear embedded within larger and spatially concentrated dissipation values. For comparison, figure \ref{fig_dissipation_raw_wiener_inst} shows the corresponding filtered image assuming a delta function as the PSF. Including the measured PSF in the Wiener filter restores the mixture fraction gradients to larger magnitude, as opposed to assuming perfect imaging. Noise evident in the latter image has, additionally, been removed as a result of the different form of the Wiener filter. The difference between the filtered and the raw scalar dissipation distribution is, also, given in figure \ref{fig_dissipation_raw_wiener_inst}. The largest correction applied to the measured image in order to arrive at the filtered result is of the order of 40\%. We can assess this level of correction by considering the apparent dissipation, \textit{e.g.} \cite{test_citation}, which quantifies the effect of noise in the true scalar dissipation value. This quantity is defined as $\left<\chi_a\right>=\frac{2\sigma_n^2}{\Delta x^2}=\left<\chi_m\right>-\left<\chi_t\right>$ (considering, \textit{e.g.}, forward differencing for the derivative), where subscripts $a, m, t$ correspond to the apparent, measured and true dissipation values, respectively, $\sigma_n^2$ is the noise variance and $\Delta x$ is the spatial discretisation. A rough estimate of this quantity for the present experimental setup results in $\sim30\%$ of the measured value, which is comparable to the actual correction imposed on the measurements. The differences between the measured and corrected images are mainly concentrated at the edges of the jet and they are typically positive, meaning that the raw image gradients have been restored by the filtering operation; within the jet body the corrections have a negative bias. This behaviour of the filtering operation results since the PSF acts to restore the large gradients, concentrated at the jet boundaries, whereas the attenuated gradients within the jet body are expected for a filtering technique like the Wiener filter, which treats the measurements in a 'global' manner, without being able to adapt to local conditions of the flow field.

\subsection{Discussion}
For any given experiment, the outcome of applying of the Wiener filter for the post processing of scalar measurements depends in an interconnected way on both the signal to noise ratio and the measurement spatial resolution. This is more clearly seen by rewriting the quantity $W(\kappa)/H(\kappa)$ appearing in Eq. \ref{eq_wiener_filtered_image} as
\[
G(\kappa) = \frac{1}{H(\kappa)}\ \frac{1}{1+S_n(\kappa)/S_s(\kappa)}
\]
where $S_s=H^2S$ is the blurred, but noise-free, spectrum (essentially the model spectrum in figure \ref{fig_wiener_model_spectra}) and the ratio $S_n/S_s$ is the inverse of the signal to noise ratio as a function of the wavenumber. At the highest wavenumbers we expect the signal to noise ratio to fall off exponentially fast, and faster than the PSF, so the very large, and noisier, wavenumbers of the signal will be attenuated. At lower wavenumbers, the form of the ratio $W(\kappa)/H(\kappa)$ depends on the relative fall-off of the 'characteristic' bandwidths of the PSF and the measurement. For example, the characteristic wavenumbers can be taken for PSF as being where the PSF starts to decrease and for the signal the wavenumber where the noise spectrum becomes important and thus where the measured spectrum diverts from the real one. If the real spectrum falls off earlier than the PSF, then the Wiener filter will roll off smoothly and will decrease the effect of the noise only. In the case when the bandwidth of the flow scales is comparable to the PSF resolution bandwidth, the Wiener filter will take a form as to initially accentuate the blurred image, before subsequently attenuating the noise components; in this situation some imaging artifacts might be created. In the case when the measurement resolution is very coarse, a lot of information will have been lost and the Wiener filter might not be the best solution to reconstruct the measured signal.

The qualitative discussion above is now made quantitative by considering the filtering process for flows exhibiting higher levels of turbulence, by varying the relative magnitudes of the flow length scales as compared to the experimental spatial resolution. We use the measured PSF so as to retain the effect of the intensifier.

\begin{figure}[htbp]
\centering
\includegraphics[scale=0.4]{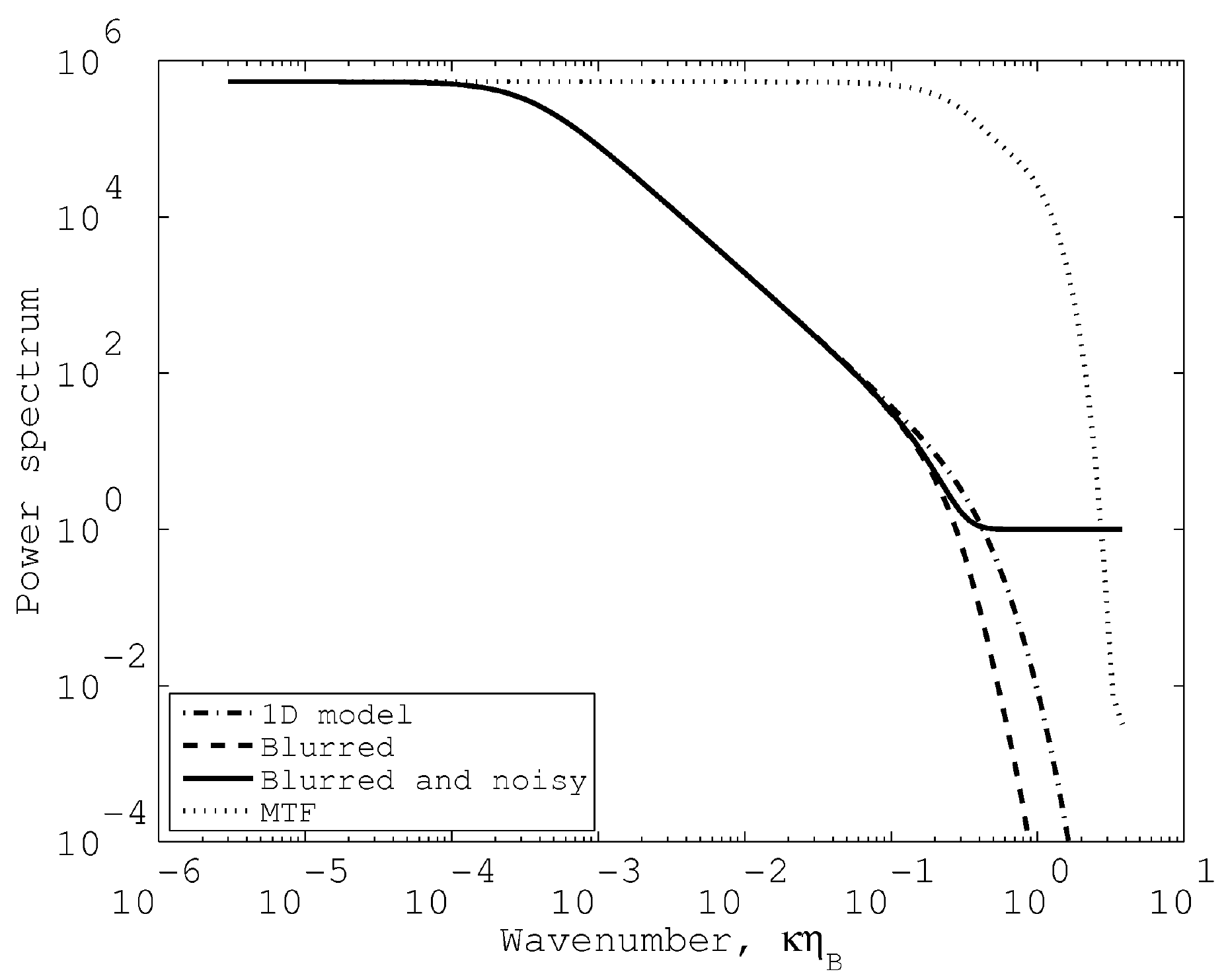}
\caption{The model 1D spectrum, its blurred version, its blurred and noisy version and the measured PSF. The Taylor-scale Reynolds number is 1000 and $\eta_B/\eta_{MTF}=1.25$.}
\label{fig_model_spectrum_study}
\end{figure}

We adapt a model for the 3D spectrum of velocity \cite{Pope_book} is adapted for use as a model for the 3D spectrum of a scalar, as in \cite{barlow_system_model}, and we obtain the model 1D spectrum as $S_{\zeta}\left(\kappa_1\right)=\int_{\kappa_1}^{\infty}\kappa^{-1}E_{\zeta}\left(\kappa\right)d\kappa$, \cite{TennekesLumley}, where $E_{\zeta}\left(\kappa\right)$ is the 3D scalar spectrum. This model 1D spectrum, the measured PSF and white noise are used in Eq. \ref{eq_image_model} to obtain the 'measured' (blurred and noisy) spectrum, having signal-to-noise ratio $\sim$100 (to calculate the SNR the spectra are integrated up to the wavenumber corresponding to the Batchelor length scale). In order to model low and high turbulent conditions, the Taylor-scale Reynolds number takes the values 100 and 1000. These quantities determine the parameters of the Wiener filter using Eq. \ref{eq_wiener_filter} and allow the calculation of the filtering errors for the scalar and its dissipation using Eqs. \ref{eq_wiener_error} and \ref{eq_wiener_mean_square_error_deriv}, respectively. As an example, figure \ref{fig_model_spectrum_study}, which is equivalent to figure \ref{fig_wiener_model_spectra},  shows the 1D modeled, 'measured' and extrapolated spectra along with the measured PSF for Taylor-scale Reynolds number 1000. The resulting Wiener filter is similar to the one presented in figure \ref{fig_wiener_filter_T08}.

\begin{figure}[htbp]
\centering
\includegraphics[scale=0.4]{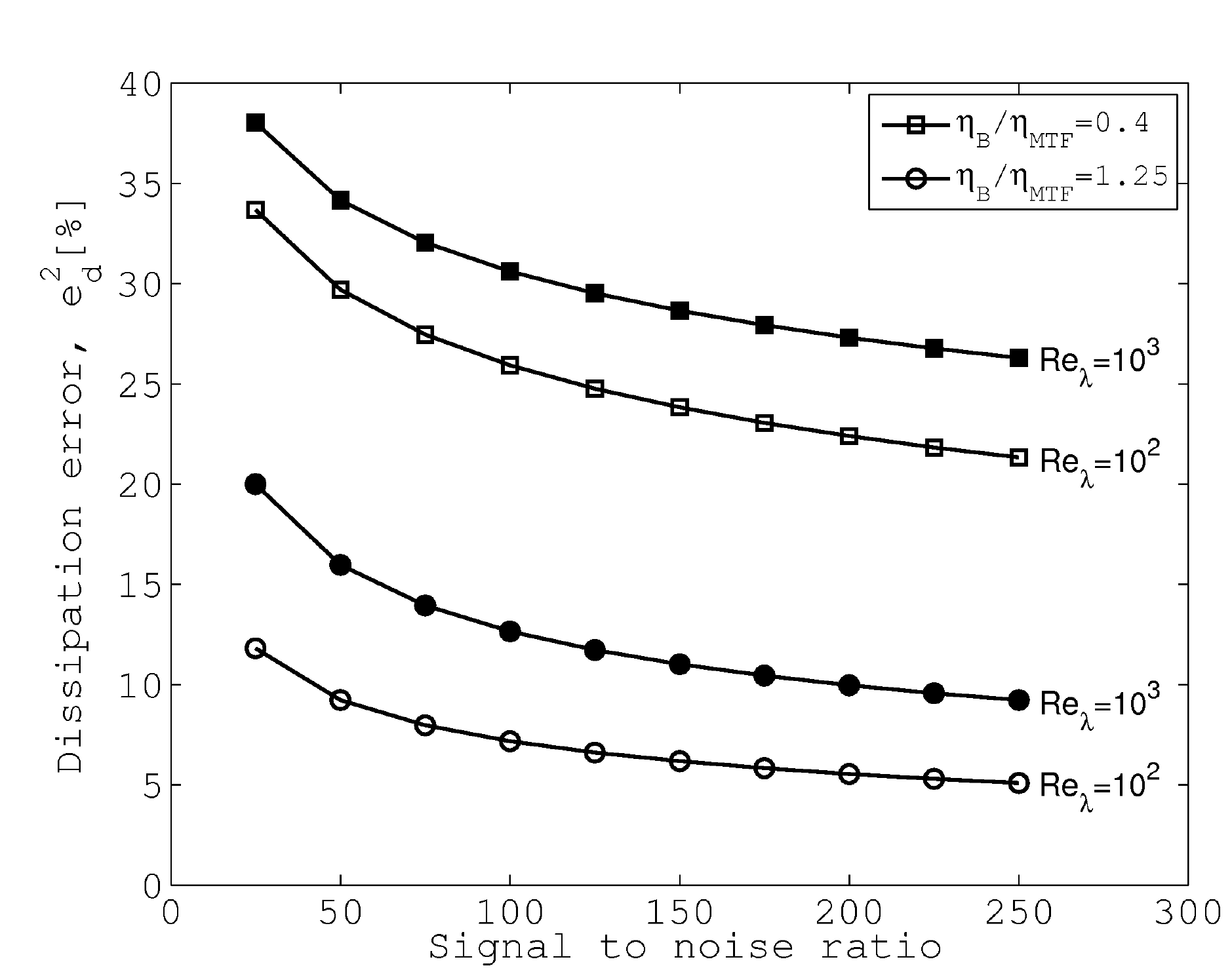}
\caption{The estimate of the error in the dissipation for different signal-to-noise ratios, for Taylor-scale Reynolds number $Re_{\lambda}=1000$ and for $\eta_B/\eta_{\mathrm{MTF}}=0.4$ and $\eta_B/\eta_{\mathrm{MTF}}=1.25$.}
\label{fig_snr_dissipation_error_study}
\end{figure}

For normalisation we use the value where the MTF is at 2\% of the peak value, $\eta_{\mathrm{\mathrm{MTF}}}=\unit[0.24]{\mathrm{mm}}$, and in figure \ref{fig_model_spectrum_study} $\eta_B/\eta_{\mathrm{MTF}}=1.25$, which corresponds to the measured values of the MTF and flow length scales. In order, however, to check the effect of the separation between the fast roll-off of the measured MTF at high wavenumbers and the Batchelor length scale in addition to the experimental noise, we also use $\eta_B/\eta_{\mathrm{MTF}}=0.4$ and a range of signal-to-noise ratios. The filtering errors for all cases are given in table \ref{table_filtering_errors_spectrum_study} and figure \ref{fig_snr_dissipation_error_study}.

\begin{table*}
\centering
\caption{The filtering errors in the estimate of the scalar and its dissipation for a blurred and noisy model 1D spectrum, with parameters the Taylor-scale Reynolds numbers, $Re_{\lambda}$, and the Batchelor length scale, for signal-to-noise ratio 100.}
\label{table_filtering_errors_spectrum_study}
\begin{tabular}{ccccc}
$Re_{\lambda}$		& $\eta_B/\eta_{\mathrm{MTF}}$				& SNR	& $e^{2}[\%]$	& $e^{2}_{d}[\%]$ \\
\hline
100				& 1.25							& 100	& 0.01		& 7.2 \\
100				& 0.4								& 100	& 0.05		& 25.9 \\
1000				& 1.25							& 100	& 5E-4		& 12.7 \\
1000				& 0.4								& 100	& 1E-3		& 30.6 \\
\end{tabular}
\end{table*}

In the cases in table \ref{table_filtering_errors_spectrum_study} the characteristic scale PSF is either larger or smaller than the Batchelor length scale. As $\eta_B$ becomes larger than $\eta_{\mathrm{MTF}}$, the error in the scalar dissipation rate decreases, for both small and large Taylor Reynolds number. So, the first conclusion is that, for flow fields described by these spectra, as long as the Batchelor length scale is well resolved by the PSF, the Wiener filter is capable of adequately filtering the measurements to produce a value for the scalar dissipation rate within acceptable limits.

In an imaging experiment, increasing the Taylor-scale Reynolds number results in a reduction in the size of the smallest flow length scales and, if one keeps the experimental setup unchanged, both resolution and noise are negatively affected; so, the measurement errors become larger. The ideal case is a simultaneous increase in spatial resolution (increasing the experiment size is a potential avenue to achieve that, wherever possible) and a proportional increase in signal power, perhaps by reducing the laser sheet size (even though this would eventually restrict the lowest resolvable wavenumber). Alternatively, one can attempt to tweak either the spatial resolution or the signal level and a common experimental situation, as the Taylor-scale Reynolds number increases, is to keep the PSF and the SNR constant. This situation is presented in table  \ref{table_filtering_errors_spectrum_study} in rows 1 and 4, where the spatial resolution reduces from $\eta_B/\eta_{\mathrm{MTF}}=1.25$ to $\eta_B/\eta_{\mathrm{MTF}}=0.4$ as the Taylor-scale Reynolds number increases from 100 to 1000. The resulting error in the estimate of the variance of the scalar decreases from $10^{-2}\%$ to $10^{-3}\%$, whereas the error in the estimate of the scalar dissipation rate increases from $7.2\%$ to $30.6\%$. The decrease in the scalar variance error results from capturing a larger part of the low-wavenumber region of the spectrum as the Batchelor length scale decreases, while the imaging field of view remains the same. In contrast, the increase of the error in the scalar dissipation is expected, given the lower spatial resolution.

Although the common experimental situation is as described above, figure \ref{fig_snr_dissipation_error_study} shows that it is more effective to expand efforts in increasing the spatial resolution rather than to improve the SNR level. Figure \ref{fig_snr_dissipation_error_study} shows the effect of the signal-to-noise ratio at $Re_{\lambda}=10^2$ and $10^3$ when $\eta_B/\eta_{\mathrm{MTF}}=0.4$ or $1.25$. As expected, the error reduces with increasing signal-to-noise ratio; increasing the signal-to-noise ratio by order of magnitude decreases the error by a third in the case of $\eta_B/\eta_{\mathrm{MTF}}=0.4$, whereas the error decreases by half when $\eta_B/\eta_{\mathrm{MTF}}=1.25$. At the same time no increase in the signal-to-noise ratio seems to be able to achieve the error afforded when the flow length scales are better resolved; \textit{e.g.} in figure \ref{fig_snr_dissipation_error_study}, the line with $\eta_B/\eta_{\mathrm{MTF}}=0.4$ and $Re_{\lambda}=10^3$ (filled squares) is always higher than the line with $\eta_B/\eta_{\mathrm{MTF}}=1.25$ at the same $Re_{\lambda}=10^3$ (filled circles). The implication is that attempting to achieve better measurement resolution at a given signal-to-noise ratio is the more profitable path to decreasing the error in the estimate of the dissipation. Further evidence of this point of view is provided by the comparison of the error variation for $Re_{\lambda}=10^2$ between $\eta_B/\eta_{\mathrm{MTF}}=0.4$ and $\eta_B/\eta_{\mathrm{MTF}}=1.25$, in figure \ref{fig_snr_dissipation_error_study}. Having a better resolution can counteract even an increase of an order of magnitude of the Taylor scale Reynolds number, irrespective of the signal-to-noise ratio.

The preceding results corroborate the earlier qualitative discussion that the relative magnitude of the measurement resolution cut-off with respect to the flow scales (\textit{e.g.} the ratio $\eta_B/\eta_{\mathrm{MTF}}$) is the determining factor in the successful application of the Wiener filter. So, for the scales in this experiment, the spatial resolution is good enough to resolve the scalar dissipation rate using the Wiener filter, but, for a flow with smaller length scales an imaging system with a narrower PSF should be used. However, the Reynolds number by itself should not be the determining factor in deciding the use of the Wiener filter in a particular situation.

%%%%%%%%%%%%%%%%%%%%%
\section{Conclusions} \label{section_conclusions}
An implementation of the Wiener filter for the measurement of mixture fraction in a starting jet flow which is in-homogeneous at the largest scales of motion is presented. An explicit calculation of the accuracy of the mixture fraction variance and of the scalar dissipation rate showed that, in the present measurements, the Wiener filter produced results that are, on average, within 2\% of the mixture fraction variance and within 20\% of the scalar dissipation rate. This accuracy estimate was, also, checked using an independent correction method and the results reinforced the applicability of the Wiener filter approach in the present measurements. We choose the Wiener filter because the error calculation gives an explicit estimate of the accuracy of the measurement of the scalar dissipation rate and because the Wiener filter can also correct the instantaneous images of the scalar dissipation rate. The results obtained from these two methods are as accurate as each other and, since the two methods are derived in different ways, we can argue that the Wiener filter is a credible approach in evaluating the scalar dissipation rate in this flow.

The analytical tractability is an advantage of the Wiener filter, as compared to other filtering methods that have been used for making scalar dissipation rate measurements. Furthermore, it does not rely on \textit{ad hoc} assumptions and is robust in the spectral modelling involved. Also, since it treats explicitly both spatial resolution and noise effects, it can point to potential improvements in either of these.

A detailed presentation of the effect of filtering in the instantaneous raw images showed that the corrections in the mixture fraction field were less than 1\% of the measured value for most times ASI and in the case of the scalar dissipation rate these were of the order of 40\%. The residual noise after the application of the filter in the mixture fraction fields was gaussian and constant at the smallest scales, which shows that the Wiener filter was capable of largely removing small scale measurement noise rather than flow information.

The procedure outlined in section \ref{section_wiener_method_noise} is, also, a way to approach the application of the Wiener filter in other flow configurations, \textit{e.g.} flows with higher Reynolds numbers or flows with combustion. In all situations the results of the filtering operation should be tested and verified against the filtering assumptions and, if possible, independently of the applied filter, similarly to the present comparison with the method of \cite{eaton_2007_dissipation_correction_piv}.

High values of the scalar dissipation rate were shown to be concentrated at the jet boundary and at random regions within the jet body. Before filtering, noise dominated the instantaneous scalar dissipation distribution, at least qualitatively, which was removed after the application of the Wiener filter. The instantaneous gradients were shown to remain after filtering and regions of low dissipation values were found, as should be expected of the instantaneous scalar dissipation distributions.

Finally, use of a model spectrum for turbulent flows with high Taylor-scale Reynolds number and the present experimental conditions, revealed that the Wiener filter produces errors similar to the errors found in the measurements. The separation between the Batchelor length scale and fall-off of the PSF and, also, the signal-to-noise ratio are the most important parameters in determining the effectiveness of the Wiener filter.

\begin{acknowledgements}
The authors acknowledge support from EPSRC grant GR/R54767/01. The automotive gas injector was donated by Keihin Corporation.
\end{acknowledgements}

\bibliographystyle{spbasic}      % basic style, author-year citations
\bibliography{Biblio}   % name your BibTeX data base

%%%%%%%%%%%%%%
%%%%%%%%%%%%%%
\end{document}